\documentclass[a4paper,11pt]{article}
\pdfoutput=1 
\usepackage{physics}

\usepackage{jheppub} 
\usepackage[T1]{fontenc} 
\usepackage{slashed}
\newcommand{\be}{\begin{equation}}
\newcommand{\ee}{\end{equation}}
\usepackage{tikz}

\newcommand{\notimplies}{%
  \mathrel{{\ooalign{\hidewidth$\not\phantom{=}$\hidewidth\cr$\implies$}}}}
\usepackage{ucs}
\usepackage{hyperref}
\usepackage[utf8x]{inputenc}
\usepackage{amsmath,bm}
\usepackage{amsfonts}
\usepackage{comment}
\usepackage{amssymb}
\usepackage{array}
\usepackage{epsfig}
\usepackage{slashed}
\usepackage{lipsum}
\usepackage{hyperref}
\hypersetup{
	colorlinks=true,
	linkcolor=blue,
	filecolor=red,
	urlcolor=blue,
	citecolor=red
}

\title{Relation between parity-even and parity-odd CFT correlation functions in three dimensions}
\author{Sachin Jain,}
\author{ Renjan Rajan John}

\affiliation{Indian Institute of Science Education and Research, Homi Bhabha Road, Pashan, Pune 411 008, India}
\emailAdd{sachin.jain@iiserpune.ac.in}
\emailAdd{renjan.john@acads.iiserpune.ac.in}

\abstract{In this paper we relate the parity-odd part of two and three point correlation functions in theories with  exactly conserved  or weakly broken higher spin symmetries to the parity-even part which can be computed from free theories. We also comment on higher point functions.

The well known connection of CFT correlation functions with de-Sitter amplitudes in one higher dimension implies a relation between  parity-even and parity-odd amplitudes calculated using non-minimal interactions such as ${\mathcal W}^3$ and ${\mathcal W}^2 {\widetilde {\mathcal W}}$.  In the flat-space limit this implies a relation between parity-even and parity-odd parts of flat-space scattering amplitudes.}

\begin{document}
	
\maketitle

\raggedbottom

\section{Introduction}
Three dimensional conformal field theory finds diverse applications in topics from cosmology to condensed matter physics  \cite{Maldacena:2011nz,Mata:2012bx,Geracie:2015drf,Karch:2016sxi,Seiberg:2016gmd,Murugan:2016zal,Minwalla:2020ysu}. They also provide interesting examples of duality. For example, the
ABJM super-conformal field theory is dual to $M$-theory on $AdS_4\times S^7$ \cite{Aharony:2008gk,Aharony:2008ug}. Another example is that of higher spin Vasiliev theories \cite{Vasiliev:1992av,Vasiliev:1995dn,Vasiliev:1999ba,Vasiliev:2003ev} which are dual to certain three-dimensional CFTs \cite{Klebanov:2002ja,Sezgin:2002rt,Giombi:2009wh}.
 It also provides interesting examples of non-supersymmetric as well as supersymmetric  dual pairs of field theory without gravity \cite{Giombi:2011kc,Aharony:2011jz,Jain:2012qi,Gur-Ari:2015pca}.
One interesting aspect of three dimensional CFT is that conformal invariance allows the  correlation functions to have  parity-odd contribution \cite{Giombi:2011rz,Costa:2011mg,Costa:2011dw}. For example, the two-point functions of spinning operators have both parity-even and parity-odd contributions. The parity-even part can be reproduced from the free theory but the parity-odd contribution can only be obtained in interacting theories  \cite{Giombi:2011kc,Aharony:2011jz,Maldacena:2011jn,Maldacena:2012sf,Giombi:2016zwa}. For three-point functions of exactly conserved currents, it was argued that in general there exist one parity-odd and two parity-even contributions \cite{Giombi:2011rz}. The two parity-even structures are related to the free bosonic or free fermionic theory. However, the parity-odd contribution again requires us to consider interacting theories such as Chern-Simons matter theories \cite{Giombi:2011kc,Aharony:2011jz,Maldacena:2011jn,Maldacena:2012sf,Giombi:2016zwa} \footnote{The parity-odd part of  $\langle J_{s} J_{s}O_{\Delta=2}\rangle$ can be obtained from the free fermionic theory in three dimensions. Similar conclusion holds for  $\langle J_{s_1} J_{s_2}O_{\Delta=2}\rangle$. Interestingly, $\langle J_{s_1} J_{s_2}O_{\Delta=2}\rangle_{\text{even}}$ can only be obtained from interacting theories.}.

The position space analysis of CFT correlation functions has attracted a great deal of attention \cite{Osborn:1993cr,Erdmenger:1996yc,Giombi:2011rz,Costa:2011mg,Costa:2011dw}. However, much less attention has been devoted to understanding correlation function in  momentum space  or in spinor-helicity variables. See \cite{Coriano:2013jba,Bzowski:2013sza,Bzowski:2017poo,Isono:2018rrb,Bzowski:2018fql,Isono:2019ihz,Isono:2019wex,Coriano:2019nkw,Coriano:2020ees,Jain:2020rmw,Jain:2021wyn,Caron-Huot:2021kjy,Jain:2021vrv} and references therein for recent progress. 
For the spinor-helicity formalism in $AdS_4$, see \cite{Nagaraj:2018nxq}. Using results of \cite{Metsaev:2018xip} helicity structures of three-point correlators of higher-spin currents in momentum space and their relation with bulk $AdS$ couplings were
discussed in \cite{Skvortsov:2018uru}. In momentum space using the light cone gauge, relations between the parity even and parity odd parts of a correlator can be observed from the results  quoted in \cite{Aharony:2012nh,GurAri:2012is,Skvortsov:2018uru} which were given in light cone coordinates for specific kinematic regions. 
In \cite{Skvortsov:2018uru}, it was shown that the parity-breaking parameter $\theta$ results from certain EM duality in the bulk.

Recently it was realised that the parity-odd contribution to correlation functions of exactly conserved currents can be obtained from the parity-even contribution in spinor-helicity variables \cite{Caron-Huot:2021kjy,Jain:2021vrv}. Using this result, it was shown in \cite{Gandhi:2021gwn}, that correlation functions in spinor-helicity variables in Chern-Simons matter theories can be written down using either free fermionic or free bosonic results. Naturally this poses the question, if this relation is special to spinor-helicity variables, or  can one find out such relations in position space and momentum space as well? The aim of this note is to write down a precise relation between the parity-even and the parity-odd contributions to a correlator in position and momentum space. This relation in particular would imply that the parity-odd part of the correlation function of conserved currents can be obtained from free theory. 
In \cite{Maldacena:2012sf}, the consistency of position space higher spin equations required a relation between the parity-odd and parity-even correlators for a few three-point correlators for specific components, but these were not explicitly checked. 
Recently in \cite{Caron-Huot:2021kjy} a conformally invariant transformation that maps parity-even correlators to parity-odd correlators was constructed and this was explicitly checked at the level of two-point  functions \footnote{We became aware of the results of \cite{Caron-Huot:2021kjy} and the similarity with some of our results when we were done with the main results of our paper.}. In this paper we make use of the splitting of correlation functions into their homogeneous and non-homogeneous terms \cite{Jain:2021qcl,Jain:2021vrv} to relate the parity-even and the parity-odd correlators in position and momentum spaces as well as in spinor-helicity variables.

We will see that for correlation functions of exactly conserved currents with spins satisfying triangle inequality, the parity-odd contribution can be obtained from the homogeneous part of the parity-even contribution. One can think of this relation as a map from parity-even correlators to parity-odd correlators. For the non-homogeneous part, there is no analogous parity-odd contribution. When the triangle inequality is violated, there is no parity-odd contribution to the correlation function of exactly conserved currents. However, for weakly broken higher-spin theories, it turns out that such correlation functions can also have a parity-odd contribution. Interestingly, when the triangle inequality is violated the only contribution to the parity-odd and the parity-even parts is the non-homogeneous contribution. In this case as well, one can find out a relation between the parity-even and the parity-odd parts of the correlation function.

Rest of the paper is structured as follows. In Section \ref{2} we discuss the parity-even and the parity-odd projection operators that we use in this paper and the relation between them. In Section \ref{3} we discuss the relation between the parity-even and the parity-odd two-point functions of conserved currents of arbitrary spin $s$. In Section \ref{4} we describe the homogeneous and non-homogeneous contributions to 3-point functions comprising spinning operators in position space, momentum space and in spinor-helicity variables. We also describe these contributions to $dS_4$  amplitudes. In Section \ref{5} we describe the relation between the parity-even and the parity-odd homogeneous  parts of three-point correlation functions of conserved currents, in spinor-helicity variables and in position and momentum spaces. To describe the relation in momentum space, we rewrite momentum space expressions using a few invariants such that the relation becomes manifest. We also extend our analysis to four-point functions. We then look at similar relations for three-point $dS_4$ amplitudes.  
In Section \ref{HSbroken1} we use higher spin equations associated to weakly broken higher spin symmetry to derive the relation between parity-odd and parity-even correlators. We conclude with a summary and future directions of study in Section \ref{9}. In Appendix \ref{mom-rev-1} we express three-point functions of conserved currents in terms of certain fundamental building blocks in momentum space. In Appendix \ref{hsap} we collect some details of higher spin equations that relate the parity-odd and the parity-even parts of correlators comprising higher spin currents.

\section{Relation between parity-even and parity-odd projection operators}
\label{2}
In this section we collect the orthogonal projectors that will be useful in the subsequent sections. We use the following parity-even projector to project spin-1 conserved currents
\begin{align}\label{pevenspin1}
    \pi^{\mu\nu}(k) = \delta^{\mu\nu}-\frac{k^{\mu}k^{\nu}}{k^2}
\end{align}
We can similarly define a parity-odd projector as \footnote{While these are explicitly orthogonal to their momentum arguments they are different from usual projectors in the sense that they do not square to themselves. However, a contraction of three of these `projectors' gives back the 'projector'.}
\begin{align}
    \chi^{\mu\nu}(k) \equiv \frac{1}{k}\epsilon^{\mu\nu k}
\end{align}
where we have used the notation $\epsilon^{\mu\nu k}\equiv\epsilon^{\mu\nu \rho}k_{\rho}$. For spin-2 conserved currents, projectors are defined such that they are orthogonal and traceless. The parity-even and parity-odd projectors in this case are given by
\begin{align}
    \Pi^{\mu\nu\rho\sigma}(k) &= \frac{1}{2}\left(\pi^{\mu\rho}(k)\pi^{\nu\sigma}(k)+\pi^{\mu\sigma}(k)\pi^{\nu\rho}(k)-\pi^{\mu\nu}(k)\pi^{\rho\sigma}(k)\right)\\[5 pt]
     \Delta^{\mu \nu \rho \sigma}(k)&=\frac 14[\epsilon^{\mu \rho k} \pi^{\nu \sigma}(k)+\epsilon^{\mu \sigma k} \pi^{\nu \rho}(k)+\epsilon^{\nu \sigma k} \pi^{\mu \rho}(k)+\epsilon^{\nu \rho k} \pi^{\mu \sigma}(k)]
    \end{align}
\subsection*{A relation between even and odd projectors}
We observe the following relations between the parity-even and the parity-odd projectors defined above. For the spin-1 case, we have
\begin{align}
\begin{split}\label{evenoddprojector}
   \frac{1}{k}\epsilon^{\mu\alpha k} \pi^{\nu}_{\alpha}(k) = \chi^{\mu\nu}(k),~~~
    \frac{1}{k}\epsilon^{\mu\alpha k} \chi^{\,\,\nu}_{\alpha}(k) = -\pi^{\mu\nu}(k)
    \end{split}
\end{align}
Similarly, for spin-2 projectors we have
\begin{align}
    \epsilon^{\alpha k(\mu}\Pi^{\nu)\rho\sigma}_{\alpha}(k)=\Delta^{\mu\nu\rho\sigma}(k)
\end{align}
where $(\mu,\nu)$ indicates symmetrisation with respect to $\mu$ and $\nu$ with a factor of $1/2$. It is also convenient to state these relations in terms of transverse, null polarization vectors which satisfy $k_i\cdot z_i =0$ and $z_i^2=0$. To do this, let us first write down the form of the projectors after contracting with these vectors
\begin{align}
\begin{split}
   z_{1\mu}z_{2\nu} \pi^{\mu\nu}(k)&=z_1 \cdot z_2\\[5 pt]
   z_{1\mu}z_{2\nu} \chi^{\mu\nu}(k)&= \frac{1}{k}\epsilon^{z_1 z_2 k}
    \end{split}
    \begin{split}
       z_{1\mu}z_{1\nu}z_{2\rho}z_{2\sigma} \Pi^{\mu\nu\rho\sigma}(k) &= (z_1 \cdot z_2)^2\\[5 pt]
       z_{1\mu}z_{1\nu}z_{2\rho}z_{2\sigma} \Delta^{\mu\nu\rho\sigma}(k) &= \epsilon^{z_1 z_2 k}(z_1 \cdot z_2)
    \end{split}
\end{align}
For the spin-$s$ case we have 
\begin{equation}
    P_s^{\text{even}}= \left(z_1\cdot z_2\right)^s,~~~~P_s^{\text{odd}}= \left(z_1\cdot z_2\right)^{s-1} \frac{1}{k}\epsilon^{z_1 z_2 k}
\end{equation}
We notice that the following substitution 
\begin{equation}\label{sub}
    z_1\cdot z_2 \rightarrow \frac{1}{k}\epsilon^{z_1 z_2 k}
\end{equation} takes the parity-even projector to the parity-odd projector.
More formally, the following operation
\begin{align}\label{opodev}
    \frac{1}{k}\epsilon^{z_1 z_2 k} \frac{1}{s} \frac{\partial}{\partial (z_1\cdot z_2)} : P_s^{\text{even}}(k)\rightarrow P_s^{\text{odd}}(k).
\end{align}
takes the parity-even projector to the parity-odd projector.

\section{Relation between parity-even and parity-odd two-point functions}
\label{3}
In this section we establish the relation between the parity-odd and the parity-even parts of two-point correlation functions in a generic 3d CFT. We will concern ourselves with only the momentum dependent  structures and ignore constants. 
We start with the 2-point function of the spin-1 current.
\subsection*{$\langle JJ\rangle$}
The parity-even and parity-odd parts of the two-point function of the spin-1 current are given by 
\begin{align}
    \langle J^{\mu}(k)J^{\nu}(-k) \rangle_{\text{even}} = \pi^{\mu\nu}(k)\,k,~~~~\langle J^{\mu}(k) J^{\nu}(-k) \rangle_{\text{odd}} =\epsilon^{\mu\nu k}
\end{align}
One can easily check that the following relations hold
\begin{align}\label{jjrel}
   \frac{1}{k} \epsilon^{\mu\alpha k}\langle J_{\alpha}(k)J^{\nu}(-k) \rangle_{\text{even}} &= \langle J^{\mu}(k) J^{\nu}(-k) \rangle_{\text{odd}}\\
   \frac{1}{k} \epsilon^{\mu\alpha k}\langle J_{\alpha}(k)J^{\nu}(-k) \rangle_{\text{odd}} &= -\langle J^{\mu}(k) J^{\nu}(-k) \rangle_{\text{even}}
\end{align}
These relations can also be interpreted as the ones between parity-odd and parity-even projectors discussed in \eqref{evenoddprojector}.

\subsection*{$\langle TT \rangle $}
The above relation can be extended to the $\langle TT \rangle$ 2-point function. The parity-odd and the parity-even 2-point functions are given by
\begin{align}
\begin{split}
   \langle T^{\mu\nu}(k)T^{\rho\sigma}(-k) \rangle_{\text{even}} &= \Pi^{\mu\nu\rho\sigma}(k)k^3
    \end{split}
    \begin{split}
           \langle T^{\mu\nu}(k)T^{\rho\sigma}(-k) \rangle_{\text{odd}} &=  \Delta^{\mu\nu\rho\sigma}(k)k^2
    \end{split}
\end{align}
In this case, we need to symmetrize as follows
\begin{align}
  \frac 12\left[\frac{1}{k}  \epsilon^{\mu \alpha k}\langle T^{\alpha \nu}(k)T^{\rho\sigma}(-k) \rangle_{\text{even}}+ (\mu \leftrightarrow \nu)\right] &= \langle T^{\mu\nu}(k)T^{\rho\sigma}(-k) \rangle_{\text{odd}}
\end{align}
This is because the right hand side is symmetric under $(\mu \leftrightarrow \nu)$ by definition. Therefore, the left hand side must also be symmetric under this exchange. 

To generalize this to higher spin currents, it is convenient to use polarization vectors. The parity-even and the parity-odd two-point functions of spin-2 currents after contracting with polarization vectors are given by
\begin{align}\label{TT}
           z_{1\mu}z_{1\nu}z_{2\rho}z_{2\sigma}\langle T^{\mu\nu}(k)T^{\rho\sigma}(-k) \rangle_{\text{even}} &= (z_1 \cdot z_2)^2k^3,\cr
    z_{1\mu}z_{1\nu}z_{2\rho}z_{2\sigma}\langle T^{\mu\nu}(k)T^{\rho\sigma}(-k) \rangle_{\text{odd}} &=  \epsilon^{z_1 z_2 k}(z_1 \cdot z_2)k^2
\end{align}
More generally
\begin{align}
          \langle z_1^s\cdot J_s(k)  z_2^s\cdot J_s(-k) \rangle_{\text{even}} &= (z_1 \cdot z_2)^{s}k^{2s-1},\cr
   \langle z_1^s\cdot J_s(k)  z_2^s\cdot J_s(-k) \rangle_{\text{odd}} &=  \epsilon^{z_1 z_2 k}(z_1 \cdot z_2)^{s-1}k^{2s-2}
\end{align} where $z_1^s$ is shorthand notation for $z_1^{\mu_1}\cdots z^{\mu_s}$.
We see that the map in \eqref{opodev} transforms the parity-even two-point function to the parity-odd two-point function.

\subsection{Relation between parity-even and parity-odd in spinor-helicity variables}
All the statements made above can be seen manifestly in spinor-helicity variables in terms of which the $2$-point function has a unique structure and the parity-even and parity-odd correlators differ only by a factor of $i$ \footnote{For details of the spinor-helicity variables that we use in this paper we refer the reader to \cite{Maldacena:2011nz}.}. 
For the spin-$s$ current we have 
\begin{align}\label{JsJstwopointfn}
\begin{split}
\langle J^{s-}(k_1) J^{s-}(k_2) \rangle &=\left(c_{J_s}+i\,c'_{J_s}\right)\frac{\langle 12 \rangle^{2s}}{2s k_2},\\[5 pt]
\langle J^{s+}(k_1) J^{s+}(k_2) \rangle &=\left(c_{J_s}-i\,c'_{J_s}\right)\frac{\langle {\bar 1}{\bar 2} \rangle^{2s}}{2s k_2},
\end{split}
\begin{split}
\langle J^{s+}(k_1) J^{s-}(k_2) \rangle &=\left(c_{J_s}+i\,c'_{J_s}\right)\frac{\langle \bar{1}2 \rangle^{2s}}{2s k_2}\\[5 pt]
\langle J^{s-}(k_1) J^{s+}(k_2) \rangle &=\left(c_{J_s}-i\,c'_{J_s}\right)\frac{\langle  1{\bar 2} \rangle^{2s}}{2s k_2}
\end{split}
\end{align}
We now turn our attention to three and  higher point functions. To understand them, let us first explain a few important aspects of correlation functions.

We also discuss the relation between  parity-even and the parity-odd two-point functions in position space in a later section.

\section{Generality : Homogeneous (h) vs Non-homogeneous (nh) parts of CFT correlators and amplitudes}
\label{4}
In this section we discuss the distinction between the homogeneous and the non-homogeneous parts of correlation functions and amplitudes. This distinction  will be useful in relating the parity-odd and parity-even parts of correlation functions.
\subsection{h vs nh in CFT correlation functions}
Any CFT correlation function can be separated into homogeneous and non-homogeneous parts, see \cite{Maldacena:2011nz} for a discussion in momentum space. In what follows we do this identification in momentum space, spinor-helicity variables and in position space.
\subsection*{Momentum space}
The non-homogeneous part of the correlation function saturates the  Ward-Takahashi (WT) identity, i.e.
\begin{align}\label{WTmom}
   k_{1\mu_1} \left\langle J^{\mu_1\cdots \mu_{s_1}}(k_1)J_{s_2}J_{s_3}\cdots J_{s_n} \right\rangle_{\bf{h}}&=0\cr
   k_{1\mu_1} \left\langle J^{\mu_1\cdots \mu_{s_1}}(k_1)J_{s_2}J_{s_3} \cdots J_{s_n} \right\rangle_{\bf{nh}}&= \rm{terms~from~the~WT ~Identity}
\end{align} 
To be more concrete, let us consider the example of $\langle J_{\mu} J_{\nu} T_{\rho \sigma}\rangle$. The correlator can be written as
\begin{align}\label{ltsp}
 \langle J_{\mu}(k_1) J_{\nu}(k_2) T_{\rho \sigma}(k_3)\rangle &=   \langle J_{\mu}(k_1) J_{\nu}(k_2) T_{\rho \sigma}(k_3)\rangle_{\textbf{Transverse}} + \langle J_{\mu}(k_1) J_{\nu}(k_2) T_{\rho \sigma}(k_3)\rangle_{\textbf{Local}}. 
\end{align}
When dotted with external momentum, the local piece reproduces the WT identity. The transverse piece can as well be split into homogeneous and non-homogeneous pieces
\begin{align}\label{nhsp}
  \langle J_{\mu}(k_1) J_{\nu}(k_2) T_{\rho \sigma}(k_3)\rangle_{\textbf{Transverse}} =  \langle J_{\mu}(k_1) J_{\nu}(k_2) T_{\rho \sigma}(k_3)\rangle_{\textbf{Trans,nh}} +\langle J_{\mu}(k_1) J_{\nu}(k_2) T_{\rho \sigma}(k_3)\rangle_{\textbf{Trans,h}}  \end{align}
Using the representation
\begin{align}\label{hmw}
 \langle J_{\mu}(k_1) J_{\nu}(k_2) T_{\rho \sigma}(k_3)\rangle &=   \langle J_{\mu}(k_1) J_{\nu}(k_2) T_{\rho \sigma}(k_3)\rangle_{\bf{nh}} + \langle J_{\mu}(k_1) J_{\nu}(k_2) T_{\rho \sigma}(k_3)\rangle_{\bf{h}}
\end{align}
and comparing with \eqref{ltsp} and \eqref{nhsp} we obtain
\begin{align}\label{hmwa}
 \langle J_{\mu}(k_1) J_{\nu}(k_2) T_{\rho \sigma}(k_3)\rangle_{\bf{h}} &=   \langle J_{\mu}(k_1) J_{\nu}(k_2) T_{\rho \sigma}(k_3)\rangle_{\textbf{Trans,h}}\nonumber\\
 \langle J_{\mu}(k_1) J_{\nu}(k_2) T_{\rho \sigma}(k_3)\rangle_{\bf{nh}} &=   \langle J_{\mu}(k_1) J_{\nu}(k_2) T_{\rho \sigma}(k_3)\rangle_{\textbf{Trans,nh}}+\langle J_{\mu}(k_1) J_{\nu}(k_2) T_{\rho \sigma}(k_3)\rangle_{\textbf{Local}}
\end{align} It can also be shown easily by explicit computation  that the homogeneous and non-homogeneous pieces can be obtained from the free bosonic and the free fermionic theory as follows
\begin{align}\label{hnh12}
  \langle J_{\mu}(k_1) J_{\nu}(k_2) T_{\rho \sigma}(k_3)\rangle_{\bf{h}} &=  \frac{1}{2}\left( \langle J_{\mu}(k_1) J_{\nu}(k_2) T_{\rho \sigma}(k_3)\rangle_{\text{FB}}- \langle J_{\mu}(k_1) J_{\nu}(k_2) T_{\rho \sigma}(k_3)\rangle_{\text{FF}}  \right) \nonumber\\
   \langle J_{\mu}(k_1) J_{\nu}(k_2) T_{\rho \sigma}(k_3)\rangle_{\bf{nh}} &=  \frac{1}{2}\left( \langle J_{\mu}(k_1) J_{\nu}(k_2) T_{\rho \sigma}(k_3)\rangle_{\text{FB}}+\langle J_{\mu}(k_1) J_{\nu}(k_2) T_{\rho \sigma}(k_3)\rangle_{\text{FF}}  \right)
\end{align}
\subsection*{Spinor-helicity variables}
It is easiest to distinguish between homogeneous and non-homogeneous contributions in spinor-helicity variables. In spinor-helicity variables, the action of the special conformal generator on the homogeneous piece gives zero whereas on the non-homogeneous piece it gives the terms that appear in the WT identity. 
It is also the case that at the level of three-point functions, homogeneous and non-homogeneous contributions have different pole structures in $E=k_1+k_2+k_3$. Generically the homogeneous contribution is always more singular. As an example, let us consider the parity-even part of the stress-tensor three-point function \cite{Bzowski:2013sza,Baumann:2020dch}  in spinor-helicity variables 
\begin{align}\label{evenTTTso}
\langle T^-T^-T^- \rangle_{\text{even}} &= \left(c_1 \frac{c_{123}}{E^6}+c_T\frac{E^3-E b_{123}-c_{123}}{c_{123}^2}\right) \langle 12 \rangle^2 \langle 23 \rangle^2 \langle 31 \rangle^2\notag\\[5 pt]
\langle T^-T^-T^+ \rangle_{\text{even}} &= c_T \frac{(E-2k_3)^2(E^3-E b_{123}-c_{123})}{E^2 c_{123}^2} \langle 12 \rangle^2 \langle 2\bar{3} \rangle^2 \langle \bar{3}1 \rangle^2
\end{align}
where $c_{123} = k_1 k_2 k_3$ and $b_{123} = (k_1 k_2+k_2 k_3+k_3 k_1)$. Note that  $c_T$ comes from the parity-even two-point function of the stress tensor \eqref{JsJstwopointfn}.
The term proportional to $c_1$ is the homogeneous contribution whereas the term proportional to  $c_T$ is the non-homogeneous contribution. It is clear  that the pole structure in $E$ of the homogeneous and non-homogeneous pieces are different. Let us also emphasize that the homogeneous piece contributes only to the $---$ and $+++$ helicity components whereas the non-homogeneous piece  contributes to all helicity components.

\subsection*{Position space}
Let us now distinguish between the homogeneous and non-homogeneous contributions in position space. We consider two examples to illustrate the distinction. 
\subsection*{$\langle J(x_1) J(x_2) O_{\Delta}(x_3) \rangle$}
The WT identity for correlation functions of the form $\langle J(x_1) J(x_2) O_{\Delta}(x_3) \rangle$ is given by
\begin{align}
 \partial_{1\mu} \langle J^{\mu}(x_1) J^{\nu}(x_2) O_{\Delta}(x_3) \rangle =0  
\end{align}
which implies that the correlator has only a homogeneous part. In terms of certain conformal structures the parity-even and the parity-odd parts of the correlator can be expressed as follows \cite{Giombi:2011rz}
\begin{align}\label{JJO11}
   \langle J_{\mu}(x_1) J_{\nu}(x_2) O_{\Delta}(x_3)\rangle_{\text{even},\bf{h}}&= \frac{1}{|x_{12}|^{2-\Delta}|x_{23}|^{\Delta}|x_{31}|^{\Delta}} \left(\Delta Q_1 Q_2+\left(4-2\Delta\right) P_3^2\right)\nonumber\\
   \langle J_{\mu}(y_1) J_{\nu}(y_2) O_{\Delta}(y_3)\rangle_{\text{odd},\bf{h}}&= \frac{1}{|x_{12}|^{2-\Delta}|x_{23}|^{\Delta}|x_{31}|^{\Delta}} S_3.
\end{align}
For details of the notation see \cite{Giombi:2011rz}.

Let us consider another example $\langle T(x_1) J(x_2) J(x_3) \rangle$. The correlator is given by \cite{Giombi:2011rz} 
\begin{align}\label{JJTpos1}
\langle T(x_1) J(x_2) J(x_3) \rangle_{\text{even}} =   \frac{1}{|x_{12}|| x_{23}||x_{31}|}&\Big[c_1 \left(P_1^2 Q_1^2 -4 P_2^2 P_3^2 - \frac{5}{2} Q_1^2 Q_2 Q_3-2 P_1 P_2 P_3 Q_1\right) \nonumber\\
&+ c_j P_1 P_2 P_3 Q_1\Big]  
\end{align} where $c_j$ is the coefficient of the two-point function of  $J_{\mu}$ fixed by the WT identity. The term proportional to $c_1$ is homogeneous and the term proportional to $c_j$ is the non-homogeneous contribution. One can always add a homogeneous piece to the non-homogeneous piece of the correlator. Doing this, one obtains
\begin{align}\label{JJTpos11}
\langle T(x_1) J(x_2) J(x_3) \rangle_{\text{even}} =   \frac{1}{|x_{12}|| x_{23}||x_{31}|}&\Big[c_1 \left(P_1^2 Q_1^2 -4 P_2^2 P_3^2 - \frac{5}{2} Q_1^2 Q_2 Q_3-2 P_1 P_2 P_3 Q_1\right) \nonumber\\
&+ c_j \left(P_1^2 Q_1^2 -4 P_2^2 P_3^2 - \frac{5}{2} Q_1^2 Q_2 Q_3+2 P_1 P_2 P_3 Q_1\right)\Big]  
\end{align}
We observe that in the representation \eqref{JJTpos11}, the homogeneous and the non-homogeneous pieces can be obtained from each other  by taking $P_3\rightarrow -P_3$. 

One can write down a general polynomial for the parity-even homogeneous and non-homogeneous contributions to correlation functions involving general conserved currents as follows
  \begin{align}\label{hpos}
      {\mathcal{F}}_{\textbf{h}}&= e^{\frac{Q_1+Q_2+Q_3}{2}} e^{P_1+P_2-P_3}\nonumber\\
      {\mathcal{F}}_{\textbf{nh}}&= e^{\frac{Q_1+Q_2+Q_3}{2}} e^{P_1+P_2+P_3}.
  \end{align}
  As is clear we have
  \begin{equation}\label{nhhin}
 P_3\rightarrow -P_3:  {\mathcal{F}}_{\textbf{h}} \rightarrow    {\mathcal{F}}_{\textbf{nh}}
  \end{equation}
It would be interesting to find out if there exists a similar relation between homogeneous and non-homogeneous contributions in momentum space.
It is useful to note the following representation of three-point correlation functions in the free bosonic and free fermionic theories
\begin{align}\label{FBFF}
{\mathcal{F}}_{\text{FB}}&={\mathcal{F}}_{\textbf{nh}}+{\mathcal{F}}_{\textbf{h}}=e^{\frac{Q_1+Q_2+Q_3}{2}} e^{P_1+P_2}\left(e^{P_3} +e^{-P_3} \right)\nonumber\\
{\mathcal{F}}_{\text{FF}}&={\mathcal{F}}_{\textbf{nh}}-{\mathcal{F}}_{\textbf{h}}=e^{\frac{Q_1+Q_2+Q_3}{2}} e^{P_1+P_2}\left(e^{P_3} -e^{-P_3} \right).
\end{align}
which precisely matches the representation given in equations 15 and 16 of \cite{Zhiboedov:2012bm}. Let us note that \eqref{FBFF} implies
\begin{align}\label{BFinterchange}
 &P_3\rightarrow -P_3:  {\mathcal{F}}_{\text{FB}} \rightarrow    {\mathcal{F}}_{\text{FB}}\nonumber\\
 &P_3\rightarrow -P_3:  {\mathcal{F}}_{\text{FF}} \rightarrow    -{\mathcal{F}}_{\text{FF}}.
  \end{align}
 We can express the homogeneous and the non-homogeneous pieces in terms of the free boson and free fermion answers as follows
 \begin{align}\label{FBFF11}
  {\mathcal{F}}_{\textbf{h}}&=\frac{1}{2}\left({\mathcal{F}}_{\text{FB}}-{\mathcal{F}}_{\text{FF}} \right)\nonumber\\
  {\mathcal{F}}_{\textbf{nh}}&=\frac{1}{2}\left({\mathcal{F}}_{\text{FB}}+{\mathcal{F}}_{\text{FF}}\right).
 \end{align}
 A simple way to understand \eqref{FBFF11} is the following
\begin{align}\label{WTmom1}
\partial_{1\mu_1} \left\langle J^{\mu_1\cdots \mu_{s_1}}(x_1)J_{s_2}J_{s_3} \right\rangle_{\bf{h}}
  &= \partial_{1\mu_1}\left( \left\langle J^{\mu_1\cdots \mu_{s_1}}(x_1)J_{s_2}J_{s_3} \right\rangle_{\bf{FB}}-\left\langle J^{\mu_1\cdots \mu_{s_1}}(x_1)J_{s_2}J_{s_3} \right\rangle_{\bf{FF}}\right)\nonumber\\
 &={ \rm{FB~ WT~Identity}}-{ \rm{ FF~WT~Identity}}\nonumber\\
 &=0
\end{align}
where in the last step we have identified the WT identity for the bosonic and fermionic theories. This identification requires us to identify the two-point function of boson and fermion. This also fixes the relative normalization. We emphasise that in this section we have considered correlators that satisfy triangle inequality \eqref{Tine}. Outside the triangle inequality all the contributions are non-homogeneous and we discuss them in the later sections.

\subsection{h vs nh in Amplitudes}
We will now distinguish between the homogeneous and non-homogeneous contributions to amplitudes. As we shall see, the gravity amplitude automatically comes in a way which separates out the homogeneous and non-homogeneous contributions. For this purpose, let us consider the $n$-point amplitude ${\mathcal M}^{\mu_{1},\mu_{2}\cdots \mu_{s_1+s_2\cdots s_n}}(k_1,k_2\cdots,k_n)$ of spinning particles with spins $s_1,s_2\cdots s_n$. The homogeneous and non-homogeneous parts of the amplitude can be defined as
\begin{align}\label{WTA}
k_i\cdot {\mathcal M}(k_1,k_2\cdots,k_n)_{\bf{h}}&=0\cr
k_i\cdot {\mathcal M}(k_1,k_2\cdots,k_n)_{\bf{nh}}&=\rm{WT~identity}
\end{align}
As an example, let us consider the three graviton $dS_4$ amplitude. It has three contributions. Two parity-even contributions come from   ${\mathcal W}^3$  and the Einstein gravity part, and a parity-odd contribution comes from ${\mathcal W}^2 {\widetilde {\mathcal W}}$. It can be easily checked that the contributions from ${\mathcal W}^3$ and ${\mathcal W}^2 {\widetilde {\mathcal W}}$ are homogeneous whereas the Einstein gravity part is non-homogeneous. Let us check this explicitly. The contribution from ${\mathcal W}^3$ is given by \cite{Giombi:2011rz}
\begin{align}\label{w3}
{\mathcal M}_{{\mathcal W}^3}&= F(1,2) F(2,3) F(3,1)\nonumber\\
F(i,j) &= \left( \epsilon_i\cdot k_j ~\epsilon_j\cdot k_i  -\epsilon_i\cdot\epsilon_j ~k_i\cdot k_j \right)
\end{align}  where $\epsilon_i$ are transverse polarization tensors.
To  analyse \eqref{WTA} we have to replace one of the polarization tensors with the momentum 
\begin{equation}
  \epsilon_i\rightarrow k_i : {\mathcal M}_{{\mathcal W}^3}\rightarrow 0.
\end{equation} which implies that the ${\mathcal W}^3$ contribution is homogeneous. It is easy to check that a similar conclusion holds for ${\mathcal W}^2 {\widetilde {\mathcal W}}$. Let us now consider the 
contribution from Einstein gravity (EG)
\begin{align}\label{EG}
 {\mathcal M}_{EG}&=  \left(k_2\cdot \epsilon_1\, \epsilon_2\cdot\epsilon_3+\text{cyclic}\right)^2 
\end{align}
The analogue of \eqref{WTA} is
\begin{equation}
  \epsilon_i\rightarrow k_i :{\mathcal M}_{EG} =\left(k_2\cdot k_1\, \epsilon_2\cdot\epsilon_3+\text{cyclic}\right) \left(k_2\cdot \epsilon_1\,\epsilon_2\cdot\epsilon_3+\text{cyclic}\right)\ne 0.
\end{equation} Thus this contribution is non-homogeneous. It is interesting to point out that in gravity, there exists a natural distinction between the homogeneous and non-homogeneous parts. In CFT, the homogeneous and non-homogeneous parts come together as we saw in the case of the free bosonic and free fermionic theories \eqref{FBFF}.

A useful gauge where amplitudes take a simple form is given by
\begin{align}
    k^{\mu} = (k, \vec{k}) \quad \epsilon^{\mu} = (0, \vec{\epsilon})
\end{align}
In this gauge the ${\mathcal M}_{{\mathcal W}^3}$ structure reduces to 
\begin{align}\label{Mw312}
    {\mathcal M}_{{\mathcal W}^3}=&[\vec{\epsilon}_1\cdot\vec{\epsilon}_2E(E-2k_3)+2\vec{\epsilon}_1\cdot \vec{k}_2\,\vec{\epsilon}_2\cdot \vec{k}_1][\vec{\epsilon}_2\cdot\vec{\epsilon}_3E(E-2k_1)+2\vec{\epsilon}_2\cdot \vec{k}_3\,\vec{\epsilon}_3\cdot \vec{k}_2]\notag\\&[\vec{\epsilon}_3\cdot\vec{\epsilon}_1E(E-2k_2)+2\vec{\epsilon}_3\cdot \vec{k}_1\,\vec{\epsilon}_1\cdot \vec{k}_2]
\end{align}
The other interaction which gives rise to non-homogeneous contributions is the Yang-Mills term \footnote {In the terminology of the recent paper \cite{Baumann:2021fxj}, interactions with minimal coupling such as Einstein gravity and Yang-Mills give rise to non-homogeneous contributions whereas interactions with non-minimal coupling  such as ${\mathcal M}_{{\mathcal W}^3}$ contribute to the homogeneous part.} whereas terms such as $F^3$, $F^2{\widetilde F}$, $\phi F^2$ and $\phi F{\widetilde F}$ contribute to homogeneous amplitudes.

\subsection{Summary of 3-point functions for exactly conserved currents} 

  Let us consider correlation functions comprising conserved currents with spins $s_1,s_2, s_3$ such that they satisfy triangle inequality
  \begin{equation}\label{Tine}
      s_{i}\le s_{j} +s_k 
  \end{equation}
  where  $i,j,k$ can be any of $1,2,3$. The most general three-point function inside the triangle can be written as the sum of parity-even free boson and free fermion contributions and a parity-odd contribution 
  \begin{align}\label{sping1}
    \left\langle J_{s_1}J_{s_2}J_{s_3} \right\rangle= c_b \left\langle J_{s_1}J_{s_2}J_{s_3} \right\rangle_{\text{FB}}+ c_f\left\langle J_{s_1}J_{s_2}J_{s_3} \right\rangle_{\text{FF}}+ c_{\text{odd}}\left\langle J_{s_1}J_{s_2}J_{s_3} \right\rangle_{\text{odd}}
  \end{align} 
 which  can also be written as 
  \begin{align}
    \left\langle J_{s_1}J_{s_2}J_{s_3} \right\rangle= c_{\bf{h}} \left\langle J_{s_1}J_{s_2}J_{s_3} \right\rangle_{\textbf{h}}+ c_{\bf{nh}}\left\langle J_{s_1}J_{s_2}J_{s_3} \right\rangle_{\textbf{nh}}+ c_{\text{odd}}\left\langle J_{s_1}J_{s_2}J_{s_3} \right\rangle_{\text{odd}}
  \end{align}
  where we used from \eqref{hnh12}
  \begin{align}\label{gnspin1}
   \left\langle J_{s_1}J_{s_2}J_{s_3} \right\rangle_{\text{FB}}&= \left\langle J_{s_1}J_{s_2}J_{s_3} \right\rangle_{\textbf{nh}}+\left\langle J_{s_1}J_{s_2}J_{s_3} \right\rangle_{\textbf{h}}  \nonumber\\
   \left\langle J_{s_1}J_{s_2}J_{s_3} \right\rangle_{\text{FF}}&= \left\langle J_{s_1}J_{s_2}J_{s_3} \right\rangle_{\textbf{nh}}-\left\langle J_{s_1}J_{s_2}J_{s_3} \right\rangle_{\textbf{h}}
  \end{align}
  Correlation functions involving a scalar operator of scaling dimension $\Delta$ can be written as
  \begin{align}\label{sso}
   \left\langle J_{s}J_{s}O_{\Delta} \right\rangle &=c_e \left\langle J_{s}J_{s}O_{\Delta} \right\rangle_{\bf{h},\text{even}}+ c_o \left\langle J_{s}J_{s}O_{\Delta} \right\rangle_{\bf{h},\text{odd}} \nonumber\\
   \left\langle J_{s} O_{\Delta} O_{\Delta} \right\rangle &= \left\langle J_{s} O_{\Delta} O_{\Delta} \right\rangle_{\bf{nh},\text{even}}
  \end{align}
  Note that there is no homogeneous contribution to $\left\langle J_{s} O_{\Delta} O_{\Delta} \right\rangle$ and no non-homogeneous\footnote{There can be a contact term that can arise in $\langle J_s J_s O\rangle$. However such terms can be suitably redefined to zero.} contribution to $\left\langle J_{s}J_{s}O_{\Delta} \right\rangle$.
  
  Let us now consider the case when the spins violate triangle inequality, i.e. say
  \begin{align}\label{Tinv}
     s_{1}>s_{2} +s_3   
  \end{align} where for simplicity we have assumed $s_1>s_2$ and $s_2\ge s_3$. In such cases when the currents are exactly conserved there is no parity-odd contribution to the correlation function and the only contributions are non-homogeneous \cite{Jain:2021whr} 
   \begin{align}
    \left\langle J_{s_1}J_{s_2}J_{s_3} \right\rangle&=  \left\langle J_{s_1}J_{s_2}J_{s_3} \right\rangle_{\bf{nh}}\cr
    &= c_b \left\langle J_{s_1}J_{s_2}J_{s_3} \right\rangle_{\text{FB}}+ c_{f}\left\langle J_{s_1}J_{s_2}J_{s_3} \right\rangle_{\text{FF}}
  \end{align}
Thus we see that the correlator is parity-even and gets only non-homogeneous contributions.
  
One can also show that for correlation functions with one scalar operator and two exactly conserved currents with unequal spins $s_1$ and $s_2$
  \begin{align}
      &\left\langle J_{s_1}J_{s_2}O_{\Delta} \right\rangle   \ne 0 ~~\rm{for ~s_1\ne s_2~\text{for}~ \Delta=1,2}\nonumber\\  
      &\left\langle J_{s_1}J_{s_2}O_{1} \right\rangle= \left\langle J_{s_1}J_{s_2}O_{1} \right\rangle_{\text{even},\bf{nh}}\nonumber\\
      &\left\langle J_{s_1}J_{s_2}O_{2} \right\rangle = \left\langle J_{s_1}J_{s_2}O_{2} \right\rangle_{\text{odd},\bf{nh}}.
  \end{align}

 \section{Relating  parity-even and odd correlation function for exactly conserved currents}
 \label{5}
 In this section we work out the relation between the parity-even and the parity-odd parts of correlation functions in spinor-helicity variables, momentum space and in position space. 
 The homogeneous part of correlation functions in spinor-helicity variables was discussed in \cite{Jain:2021vrv}. In position space it is given by \eqref{FBFF11}. In \cite{Jain:2021vrv}, momentum space expressions were also written down. However, to construct a map from parity-even to parity-odd, it is convenient to rewrite the expressions in a slightly different way which makes the map manifest. We will start our analysis using spinor-helicity variables.
 \subsection{Relation in spinor-helicity variables} 
 Consider a general correlator of the form $\langle J_{s_1}J_{s_2}J_{s_3} \rangle$ where $s_1 \geq s_2 \geq s_3$ and $s_2+s_3 > s_1$. In spinor-helicity variables, the homogeneous part of the correlator is given by
 \begin{align}\label{s1s2s3sh}
     \langle J_{s_1}^- J_{s_2}^-J_{s_3}^- \rangle_{\mathbf{h}} &= \left(c_{\text{even}}+ i c_{\text{odd}}\right)\frac{k_1^{s_1-1}k_2^{s_2-1}k_3^{s_3-1}}{E^{s_1+s_2+s_3}}\langle 12 \rangle^{s_1+s_2-s_3}\langle 23 \rangle^{s_2+s_3-s_1}\langle 31 \rangle^{s_1+s_3-s_2}
 \end{align}
 The only other non-zero contribution to the homogeneous piece comes from the $+++$ helicity component which can be obtained by complex conjugating\eqref{s1s2s3sh}. The other helicity components have only non-homogeneous contribution \cite{Jain:2021vrv}. 
 It is interesting to note that the parity-even and the parity-odd parts of the homogeneous part of the correlation function  are the same up to factors of $i$.
 To avoid clutter, we introduce the following notation  
\begin{align}
      a&\equiv \frac{1}{2}(s_1+s_2-s_3),\quad
        b\equiv \frac{1}{2}(s_2+s_3-s_1),\quad c\equiv \frac{1}{2}(s_1+s_3-s_2)\nonumber\\[5pt]
        S&\equiv s_1+s_2+s_3
  \end{align}
 \subsection{Relation in momentum space} 
The homogeneous part of any $3$-point function of the form $\langle J_{s_1} J_{s_2}J_{s_3} \rangle$ can be written in terms of a finite number of  building blocks as was observed in \cite{Jain:2021vrv}.
In \cite{Jain:2021vrv} we had introduced
\begin{align}
\label{momentumspaceblocks}
    &Q_{12} = \frac{1}{E^{2}}\left[2\left(\vec{z}_{1} \cdot \vec{k}_{2}\right)\left(\vec{z}_{2} \cdot \vec{k}_{1}\right)+E\left(E-2 k_{3}\right) \vec{z}_{1} \cdot \vec{z}_{2}\right]\nonumber\\
    &S_{12} = \frac{2}{E^{2}}\left[k_{1} \epsilon^{k_{2} z_{1} z_{2}}-k_{2} \epsilon^{k_{1} z_{1} z_{2}}\right]\nonumber\\
   &P_{123}=\frac{1}{E^{3}}\left[2\left(\vec{z}_{1} \cdot \vec{k}_{2}\right)\left(\vec{z}_{2} \cdot \vec{k}_{3}\right)\left(\vec{z}_{3} \cdot \vec{k}_{1}\right)+E\left(k_{3}\left(\vec{z}_{1} \cdot \vec{z}_{2}\right)\left(\vec{z}_{3} \cdot \vec{k}_{1}\right)+\text { cyclic }\right)\right]\nonumber\\
   &R_{123} = \frac{1}{E^3}\left[\left\{(\vec{k}_1 \cdot \vec{z}_3)\left(\epsilon^{k_3 z_1 z_2}k_1-\epsilon^{k_1 z_1 z_2}k_3\right)+(\vec{k}_3 \cdot \vec{z}_2)\left(\epsilon^{k_1 z_1 z_3}k_2-\epsilon^{k_2 z_1 z_3}k_1\right)\right.\right.\nonumber\\[5 pt]
&\hspace{1.5cm}\left.\left.-(\vec{z}_2 \cdot \vec{z}_3)\epsilon^{k_1 k_2 z_1}E+\frac{k_1}{2} \epsilon^{z_1 z_2 z_3}E(E-2k_1)\right\}+\text{cyclic perm}\right]
\end{align}
Let us split correlators into two families, one satisfying $s_1+s_2+s_3=\text{even}$ and the other satisfying $s_1+s_2+s_3=\text{odd}$.
\subsection*{$s_1+s_2+s_3=2n\;\;(n \in \mathbb{Z})$}
The parity-even and the parity-odd homogeneous pieces of correlators involving a scalar operator are given by 
     \begin{align}\label{odevnans1}
   \langle J_s J_s O_2\rangle_{\text{even}} &= \frac{b_{12}^{2s-2}}{E^{2s-2}}Q_{12} \left(z_1\cdot z_2\right)^{s-1}\nonumber\\
    \langle J_s J_s O_2\rangle_{\text{odd}} 
    &=\frac{b_{12}^{2s-2}
     }{E^{2s-2}}S_{12} \left(z_1\cdot z_2\right)^{s-1}.
     \end{align}
Correlation functions involving all spinning operators are given by      %
     \begin{align}\label{odevnans2}
      \langle J_{s_1}J_{s_2}J_{s_3} \rangle_{\text{even}} 
     &= \frac{k_1^{s_1-1}k_2^{s_2-1}k_3^{s_3-1}}{E^{S-4}}(z_1 \cdot z_3)^{c-1}(z_2 \cdot z_3)^{b-1}(z_1 \cdot z_2)^{a} Q_{13} Q_{23}\nonumber\\
     \langle J_{s_1}J_{s_2}J_{s_3} \rangle_{\text{odd}} 
     &= \frac{k_1^{s_1-1}k_2^{s_2-1}k_3^{s_3-1}}{E^{S-4}}(z_1 \cdot z_3)^{c-1}(z_2 \cdot z_3)^{b-1}(z_1 \cdot z_2)^{a-1}\epsilon^{k_1 z_1 z_2} Q_{13} Q_{23}\nonumber\\
     &= \frac{k_1^{s_1-1}k_2^{s_2-1}k_3^{s_3-1}}{E^{S-4}}(z_1 \cdot z_3)^{c-1}(z_2 \cdot z_3)^{b-1}(z_1 \cdot z_2)^{a} S_{13} Q_{23}
     \end{align}
 Let us note that the results presented in \eqref{odevnans1} and \eqref{odevnans2} are significantly simpler than what appears in  \cite{Jain:2021vrv}.  To reach these results we have made repeated use of  degeneracy and Schouten identities. The details are worked out in  Appendix \ref{mom-rev-1}. Results in \eqref{odevnans1} and \eqref{odevnans2} can be symmetrized appropriately. However, it can be shown that all such symmetrized terms are related by  degeneracy and Schouten identities to the expressions in \eqref{odevnans1} and \eqref{odevnans2}. 
 
Let us now consider the second family of correlators in which $s_1+s_2+s_3=\text{odd}$.

\subsection*{$s_1+s_2+s_3=2n+1\;\;(n \in \mathbb{Z})$}
When $s_1+s_2+s_3=\text{odd}$ for general spins $s_1,s_2,s_3$, the correlator can be written as
 \begin{align}
  &\langle J_{s_1} J_{s_2} J_{s_3}\rangle_{\text{even},\bf{h}} = \frac{k_1^{2(s_1-1)}k_2^{2(s_2-1)}k_3^{2(s_3-1)}}{E^{S-3}} P_{123} \left(z_1\cdot z_2\right)^{a-1/2} \left(z_1\cdot z_3\right)^{c-1/2} \left(z_2\cdot z_3 \right)^{b-1/2}\\
   &\langle J_{s_1} J_{s_2} J_{s_3}\rangle_{\text{odd},\bf{h}}= \frac{k_1^{2(s_1-1)}k_2^{2(s_2-1)}k_3^{2(s_3-1)}}{E^{S-3}} R_{123} \left(z_1\cdot z_2\right)^{a-1/2} \left(z_1\cdot z_3\right)^{c-1/2} \left(z_2\cdot z_3 \right)^{b-1/2}  
 \end{align}
 
 \subsection*{Relating parity-even to parity-odd}
 From their expressions in \eqref{momentumspaceblocks} one can check that 
 \begin{align}\label{oddab}
  \frac{1}{k_1} \epsilon^{k_1 z_1 z_2}  P_{123}= -k_1 k_2\,z_1\cdot z_2\,    R_{123}
 \end{align}
 Furthermore $R_{123}$ and $P_{123}$ are related by
 \begin{align}\label{oddcd}
   \frac{1}{k_i}\epsilon^{\mu k_i z_i}\frac{\partial}{\partial z_i^{\mu}} P_{123}=  R_{123}.
 \end{align}
 We can see that \eqref{oddab} and \eqref{oddcd} map parity-even correlators to parity-odd correlators.
We see the following simple relation between the parity-even and the parity-odd results
\begin{align}\label{opodev11}
    \frac{1}{k_1}\epsilon^{z_1 z_2 k_1}  \frac{\partial}{\partial (z_1\cdot z_2)} : {\rm{even_{\bf{h}}}}\rightarrow {\rm{odd}}.
\end{align} 
Let us note that for $s_1\ne 0, s_2\ne 0, s_3\ne 0$ we could have chosen any $z_i\cdot z_j$ and replaced it with either $ \frac{1}{k_i}\epsilon^{z_i z_j k_i}$ or $\frac{1}{k_j}\epsilon^{z_i z_j k_j}$ and it would have given us the map \eqref{opodev11}. We also note that the map in \eqref{opodev11} is exactly the same as the one observed at the level of two-point functions in \eqref{opodev}. 

Using Todorov operator one can rewrite the relation in \eqref{opodev11} as
 \begin{align}\label{s1s2s3evenodd}
    \langle J_{s_1}J_{s_2}J_{s_3} \rangle_{\text{even},\bf{h}} \mapsto  \frac{1}{k_1}\epsilon_{\alpha}^{\,\,k_1 (\mu_1}\langle J_{s_1}^{\mu_2 \cdots \mu_{s_1})\alpha}J_{s_2}J_{s_3}\rangle_{\text{even},\bf{h}}
 \end{align}
 This maps the parity-even homogeneous contribution in momentum space to the parity-odd contribution 
 \begin{align}
      \langle J_{s_1}J_{s_2}J_{s_3} \rangle_{\text{odd},\bf{h}} =  \frac{1}{k_1}\epsilon_{\alpha}^{\,\,k_1 (\mu_1}\langle J_{s_1}^{\mu_2 \cdots \mu_{s_1})\alpha}J_{s_2}J_{s_3}\rangle_{\text{even},\bf{h}}
 \end{align}
 One can check that this reduces to the relation one has in spinor-helicity variables.
 
\subsection*{Summary of map}
Using all the results discussed above one can show the following in general \footnote{Let us now consider the above relation in the light-cone coordinates $ds^2=2dx^+dx^-+dx_3^2$. Let us also consider momenta that have only non-zero component along the third direction. The relation then takes the following form 
  \begin{align}
  \text{sgn}\,(k_{1,3})\langle J^{- - \cdots -}(k_1)J^{-\cdots -}(k_2)J^{-\cdots-}(k_3) \rangle_{\text{\text{even},\bf{h}}} =\langle J^{- \cdots -}(k_1)J^{- \cdots -}(k_2)J^{- \cdots -}(k_3)\rangle_{\text{odd}}
  \end{align}
}
\begin{align}\label{3ptmapaf}
   \frac{1}{k_1}\epsilon^{\mu_1 \alpha k_1}\langle J_{\alpha}^{ \mu_2 \cdots \mu_{s_1}}(k_1)J^{\nu_1 \cdots \nu_{s_2}}(k_2)J^{\rho_1 \cdots \rho_{s_3}}(k_3) \rangle_{\text{\text{even},\bf{h}}} +& (\mu_1 \leftrightarrow \mu_2)+\cdots+(\mu_1 \leftrightarrow \mu_{s_1})\nonumber\\[5 pt]
  &\hspace{-3cm}=\langle J^{\mu_1 \cdots \mu_{s_1}}(k_1)J^{\nu_1 \cdots \nu_{s_2}}(k_2)J^{\rho_1 \cdots \rho_{s_3}}(k_3)\rangle_{\text{odd}}\nonumber\\[5 pt] 
  -\frac{1}{k_1}\epsilon^{\mu_1 \alpha k_1}\langle J_{\alpha}^{ \mu_2 \cdots \mu_{s_1}}(k_1)J^{\nu_1 \cdots \nu_{s_2}}(k_2)J^{\rho_1 \cdots \rho_{s_3}}(k_3) \rangle_{\text{odd}} +& (\mu_1 \leftrightarrow \mu_2)+\cdots+(\mu_1 \leftrightarrow \mu_{s_1})\nonumber\\[5 pt]
  &\hspace{-3cm}=\langle J^{\mu_1 \cdots \mu_{s_1}}(k_1)J^{\nu_1 \cdots \nu_{s_2}}(k_2)J^{\rho_1 \cdots \rho_{s_3}}(k_3)\rangle_{\text{\text{even},\bf{h}}}.
  \end{align}
   In the above we used the first spin for mapping. However note that the mapping works out exactly the same way for the second and the third spins as well.
   
 \subsection{Relation in position space}
 \label{sectionpositionspace}
The relation between the parity-even and the parity-odd parts of the correlation function in momentum space can easily be Fourier transformed to get the relations in position space. One can define a kernel which maps the parity-even contribution to the parity-odd contribution. The inverse kernel also exists which maps the parity-odd contribution to the parity-even contribution.

Let us start with the two-point function of the spin-1 current 
\begin{align}
 \langle J_{\mu}(x)J_{\nu}(y)\rangle_{\text{even}}&=\left(-\delta_{\mu\nu}+\frac{2(x-y)_{\mu}(x-y)_{\nu}}{(x-y)^2}\right)\frac{1}{(x-y)^4} \nonumber\\
 \langle J_{\mu}(x)J_{\nu}(y)\rangle_{\text{odd}}&= \epsilon_{\mu\nu\alpha}\partial^{\alpha}\delta^3(x-y).
\end{align}
The parity-odd piece is a contact term. This is true for the parity-odd part of the two-point function of any spin-$s$ conserved current.

For the spin-1 current it is easy to check that the parity-odd and the parity-even parts are related as 
\begin{align}
    &\langle J_{\mu}(x) J_{\nu}(y)\rangle_{\text{odd}} =\epsilon_{\mu\sigma\alpha}  \int \frac{d^3 x_1}{|x-x_1|^2}~~\partial_{x_1}^\sigma  \langle J^{\alpha}(x_1) J_{\nu}(y)\rangle_{\text{even}}
\end{align}
where $\partial_{x_1}^\sigma=\frac{\partial}{\partial x_{1,\sigma}}$.

\subsection*{Three-point function}
We now turn our attention to three-point functions. A Fourier transform of the relation in momentum space gives
\begin{align}\label{oeps}
    \langle J^{\mu_1\mu_2\cdots \mu_{s_1}}(y_1) J_{s_2}(y_2) J_{s_3}(y_3)\rangle_{\text{odd}} = \epsilon^{(\mu_1}_{\sigma\alpha}  \int \frac{d^3 x_1}{|y_1-x_1|^2}~~\partial_{x_1}^\sigma  \langle J^{\alpha \mu_2\cdots\mu_{s_1})}(x_1) J_{s_2}(y_2) J_{s_3}(y_3)\rangle_{\text{even},\bf{h}}
\end{align}
where we have suppressed indices of the second and third operators. 
There are other useful representations as well such as
\begin{align}
 \epsilon^{(\mu_1}_{\sigma\alpha} \partial_{y_1}^\sigma \langle J^{\alpha\mu_2\cdots \mu_{s_1})}(y_1) J_{s_2}(y_2) J_{s_3}(y_3)\rangle_{\text{odd}} =   -\int \frac{d^3 x_1}{|y_1-x_1|^4}~~  \langle J^{\mu_1 \mu_2\cdots\mu_{s_1}}(x_1) J_{s_2}(y_2) J_{s_3}(y_3)\rangle_{\text{even},\bf{h}} 
\end{align}
It is possible to invert the relation and express the parity-even result in terms of the parity-odd result as follows
\begin{align}
    \langle J^{\mu_1\mu_2\cdots \mu_{s_1}}(y_1) J_{s_2}(y_2) J_{s_3}(y_3)\rangle_{\text{even},\bf{h}} = -\epsilon^{(\mu_1}_{\sigma\alpha}  \int \frac{d^3 x_1}{|y_1-x_1|^2}~~\partial_{x_1}^\sigma  \langle J^{\alpha \mu_2\cdots\mu_{s_1})}(x_1) J_{s_2}(y_2) J_{s_3}(y_3)\rangle_{\text{odd}}
\end{align}
which can also be rewritten as
\begin{align}
 \epsilon^{(\mu_1}_{\sigma\alpha} \partial_{y_1}^\sigma \langle J^{\alpha\mu_2\cdots \mu_{s_1})}(y_1) J_{s_2}(y_2) J_{s_3}(y_3)\rangle_{\text{even},\bf{h}} =   \int \frac{d^3 x_1}{|y_1-x_1|^4}~~  \langle J^{\mu_1 \mu_2\cdots\mu_{s_1}}(x_1) J_{s_2}(y_2) J_{s_3}(y_3)\rangle_{\text{odd}}. 
\end{align}
Using \eqref{hpos}, we can symbolically write 
\begin{align}
\label{evenoddposspace}
    \langle J^{\mu_1\mu_2\cdots \mu_{s_1}}(y_1) J_{s_2}^{\nu_1 \cdots \nu_{s_2}}(y_2) J_{s_3}^{\gamma_1 \cdots \gamma_{s_3}}(y_3)\rangle_{\text{odd}} &= \epsilon^{(\mu_1\sigma\alpha}  \int \frac{d^3 x_1}{|y_1-x_1|^2}~~\partial_{x_1}^\sigma  {\mathcal F_{\bf{h}}}^{\alpha \mu_2\cdots\mu_{s_1}) \nu_1 \cdots \nu_{s_2} \gamma_1 \cdots \gamma_{s_3}}
\end{align}
Let us denote the kernel by ${\mathcal K}$. We then have
\begin{align}
  \langle J_{s_1} J_{s_2} J_{s_3} \rangle_{\text{odd}} 
  &=  {\mathcal K} \left[{\mathcal F}_{\textbf{h}}\right]\nonumber\\
  &=   {\mathcal K} \left[e^{\frac{Q_1+Q_2+Q_3}{2}} e^{P_1+P_2+P_3}\right]\nonumber\\
 &= \frac{1}{2} {\mathcal K}\left[\langle J_{s_1} J_{s_2} J_{s_3} \rangle_{\text{FF}}-\langle J_{s_1} J_{s_2} J_{s_3} \rangle_{\text{FB}} \right].
\end{align}
We shall derive this relation in section \ref{HSbroken1}. 
As a concrete  example one can consider $\langle J_{\mu} J_{\nu} O_{\Delta}\rangle$ given in \eqref{JJO11}. Using the star triangle relation \footnote{The star triangle relation is given by
\begin{equation}
    \int d^3x\frac{1}{|x-y_1|^{2\Delta_1}|x-y_2|^{2\Delta_2}|x-y_3|^{2\Delta_3}}\propto\frac{1}{y_{12}^{3-2\Delta_3}y_{13}^{3-2\Delta_2}y_{23}^{3-2\Delta_1}}
\end{equation} where $\sum_{i=1}^3\Delta_i=3$.},
it is straightforward  to show that 
\begin{align}
    \langle J^{\mu}(x_1) J^{\nu}(x_2) O_{\Delta}(x_3)\rangle_{\text{odd}} = \epsilon^{\mu\sigma\alpha}  \int \frac{d^3 y_1}{|y_1-x_1|^2}~~\partial_{x_1}^\sigma  \langle J^{\alpha}(y_1) J^{\nu}(x_2) O_{\Delta}(x_3)\rangle_{\text{even}}.
\end{align}

\subsection{Relation between parity-even and parity-odd three-point $dS_4$ amplitude}
It is known that CFT correlation functions and $dS_4$ amplitudes are related to each other\cite{Maldacena:2011nz}. This readily implies that the parity-even and the parity-odd parts of the $dS_4$ amplitudes are related just as they are in CFT correlators. To be explicit in the computations, we will consider the three-point graviton amplitude. The parity-even parts of the correlator are given by \eqref{w3} and \eqref{EG}. The parity-odd part of the amplitude is 
\begin{align}
  {  \mathcal M}_{{\mathcal W}^2 {\mathcal {\widetilde W}}}=&\left[2(\vec{z}_1\cdot\vec{k}_2 \vec{z}_2\cdot\vec{k}_3\vec{z}_3\cdot\vec{k}_1)+E\left( (\vec{z}_1\cdot\vec{z}_2\vec{z}_3\cdot\vec{k}_1)k_3+\text{cyclic}\right)\right]\notag\\&\bigg[(\vec{z}_2\cdot\vec{k}_3)\left(\epsilon^{k_3z_1z_2}k_1+\epsilon^{k_1z_1z_3}(E-k_3)\right)+(\vec{k}_1\cdot\vec{z}_3)\left(\epsilon^{k_2z_1z_2}k_1+\epsilon^{k_1z_1z_2}(E-k_2)\right)\notag\\
  &-E(\vec{z}_2\cdot\vec{z}_3)\epsilon^{k_1k_2z_1}+\frac{1}{2}k_1E(E-2k_1)\epsilon^{z_1z_2z_3}+\text{cyclic perm.}\bigg]
\end{align}
We choose a gauge in which $k^\mu=(|\vec k|,\vec{k}_i)$ and $z^{\mu}=(0,\vec z)$. It is easy to check that the contributions from ${\mathcal W}^3$ and ${\mathcal W}^2 {\mathcal {\widetilde W}}$ are related to each other by
\begin{align}
    \frac{1}{k_i}\epsilon^{\mu k_i z_i}\frac{\partial}{\partial z_i^{\mu}}\mathcal{M}_{W^3} \propto \mathcal{M}_{W^2\widetilde{W}}
\end{align}
Similar relations hold between contributions from $\phi W\widetilde W$ and $\phi W^2$, $\phi F\widetilde F$ and $\phi F^2$, and  $F^2\widetilde F$ and $F^3$. Similar statements can be made for interactions involving higher-spins. We also notice that there is no such relation involving the amplitude given by Einstein gravity part \eqref{EG}.

\subsection{Four-point correlator}\label{6}
Very few results exist for four-point functions in momentum space \cite{Li:2019twz, Kalloor:2019xjb,Silva:2021ece}. For details on scalar four-point correlators see \cite{Bedhotiya:2015uga,Bzowski:2019kwd,Bzowski:2020kfw,Turiaci:2018nua,Yacoby:2018yvy}.

In this section we first argue abstractly that if there is a parity-even homogeneous solution to the conformal Ward identity then there also exists a parity-odd homogeneous solution. To show this let us consider the simplest example of $\langle T O O O\rangle$. 

\subsubsection{$\langle T O O O\rangle_{\text{even}}$}
The ansatz for the transverse-traceless part of the parity-even correlator is given by
\begin{align}\label{anstz1}
   \langle T^{\mu\nu}(k_1)O(k_2)O(k_3)O(k_4) \rangle_{\text{even}} &= \Pi^{\mu\nu}_{\alpha\beta}(k_1)\left(A_1 k_2^{\alpha}k_2^{\beta}+A_2 k_3^{\alpha}k_3^{\beta}+A_3k_4^{\alpha}k_4^{\beta}\right)
\end{align}
The WT identity for the correlator is \cite{Coriano:2019nkw}
\begin{align}
    k_{1\mu}\langle T^{\mu\nu}(k_1)O(k_2)O(k_3)O(k_4) \rangle_{\text{even}}
    =& -k_2^{\nu}\langle O(k_3+k_3)O(k_3)O(k_4) \rangle -k_3^{\nu}\langle O(k_2+k_3)O(k_2)O(k_4) \rangle\nonumber\\[5 pt]
   &-k_4^{\nu}\langle O(k_2+k_3)O(k_2)O(k_3) \rangle
\end{align}
We separate the correlation function into homogeneous and non-homogeneous parts 
\begin{align}\label{hnhsp1}
    \langle TOOO \rangle = \langle TOOO \rangle_{\bf{h},\text{even}} +  \langle TOOO \rangle_{\bf{nh},\text{even}}.
\end{align}
Let us now convert the ansatz in \eqref{anstz1} to spinor-helicity variables
\begin{align}\label{toooevenansatzsph}
   \langle T^-(k_1)O(k_2)O(k_3)O(k_4) \rangle_{\text{even}} &= A_1 \langle 12 \rangle^2 \langle \bar{2}1 \rangle^2+A_2 \langle 13 \rangle^2\langle \bar{3}1\rangle^2+A_3 \langle 14 \rangle^2 \langle \bar{4}1\rangle^2.
\end{align}
The positive helicity component is obtained by complex conjugation.

The conformal Ward identity in spinor-helicity variables is given by
\begin{align}\label{cnfwt1}
    \widetilde{K}^{\kappa}\left\langle \frac{T^-}{k_1} OOO \right\rangle &= 12\frac{z_1^{-\kappa}}{k_1^3}z_{\left(1\mu\right.}^{-}k_{\left.1\nu\right)}\langle T^{\mu\nu}OOO \rangle 
\end{align}
where $z_1^{-\kappa}=\frac{(\sigma^{\kappa})^{\alpha\beta}\lambda_{1\alpha}\lambda_{1\beta}}{2k_1}$. Plugging \eqref{hnhsp1} in \eqref{cnfwt1} we see that the homogeneous piece satisfies
\begin{align}\label{HToo}
     \widetilde{K}^{\kappa}\left\langle \frac{T^-}{k_1} OOO \right\rangle_{\bf{h},\text{even}} &= 0
\end{align}
In general it is not known how to solve this equation in terms of an arbitrary function of conformal cross-ratios. However, for our purpose, it is sufficient to know just the equation \eqref{HToo}. 

\subsubsection{$\langle T O O O\rangle_{\text{odd}}$}
Let us now write down ansatz for the parity-odd contribution.
It can be written as
\begin{align}\label{anstz1od}
   \langle T^{\mu\nu}(k_1)O(k_2)O(k_3)O(k_4) \rangle_{\text{odd}} &= \Delta^{\mu\nu}_{\alpha\beta}(k_1)\left(B_1 k_2^{\alpha}k_2^{\beta}+B_2 k_3^{\alpha}k_3^{\beta}+B_3 k_4^{\alpha}k_4^{\beta}\right)
\end{align}
The WT identity for the parity-odd part is trivial 
\begin{align}
    k_{1\mu}\langle T^{\mu\nu}(k_1)O(k_2)O(k_3)O(k_4) \rangle_{\text{odd}}
    =0.
\end{align}
Thus we see that the parity-odd part has only homogeneous contribution 
\begin{align}\label{hnhsp1od}
    \langle TOOO \rangle_{\text{odd}} = \langle TOOO \rangle_{\bf{h},\text{odd}}.
\end{align}
Transforming the ansatz \eqref{anstz1od} to spinor-helicity variables we obtain
\begin{align}\label{spinod1}
    \langle T^- OOO \rangle_{\text{odd}} &= B_1 \langle 12 \rangle^2 \langle \bar{2}1 \rangle^2+B_2 \langle 13 \rangle^2\langle \bar{3}1\rangle^2+B_3 \langle 14 \rangle^2 \langle \bar{4}1\rangle^2
\end{align}
Let us note that even though the parity-odd ansatz \eqref{anstz1od} looks completely different from the parity-even ansatz\eqref{anstz1}, in spinor-helicity variables they become identical.
The conformal Ward identity is again given by
\begin{align}\label{HTood}
     \widetilde{K}^{\kappa}\left\langle \frac{T^-}{k_1} OOO \right\rangle_{\bf{h},\text{odd}} &= 0.
\end{align}
It is easy to show that if $\{A_1,A_2, A_3\}$ is a solution to \eqref{HToo}, one of the solutions of \eqref{HTood} is $\{B_1,B_2, B_3\}$ where
\begin{equation}
    B_1= A_1,~~~B_2=A_2,~~~B_3=A_3.
\end{equation}
As in the case of three-point functions, we find a map between the parity-even and the parity-odd homogeneous parts of the correlator as follows
\begin{align}\label{evenboddtr}
 \frac{1}{k_1}\epsilon^{\;\;k_1 \left(\mu\right.}_{\gamma}\langle T^{\left.\nu\right)\gamma}OOO \rangle_{\text{even}}&=\frac{1}{k_1}\epsilon^{\;\;k_1 \left(\mu\right.}_{\gamma} \Pi^{\nu)\gamma}_{\alpha\beta}(k_1)\left(A_1 k_2^{\alpha}k_2^{\beta}+A_2 k_3^{\alpha}k_3^{\beta}+A_3k_4^{\alpha}k_4^{\beta}\right)\nonumber\\
 &= \Delta^{\mu \nu}_{\alpha\beta}(k_1)\left(A_1 k_2^{\alpha}k_2^{\beta}+A_2 k_3^{\alpha}k_3^{\beta}+A_3k_4^{\alpha}k_4^{\beta}\right)\nonumber\\
 &=\langle T^{\mu\nu} OOO \rangle_{\text{odd}}
\end{align}
One can easily generalize this discussion to correlation functions involving more than one spinning operator as well as to arbitrary $n$-point functions. We conclude that for an arbitrary $n$-point function, we can construct a parity-odd homogeneous solution using a parity-even homogeneous solution by doing the epsilon-transformation we discussed above.

\section{A derivation using weakly broken higher spin symmetry}\label{HSbroken1}
In this section we make use of weakly broken higher spin symmetry to derive relations  presented in the previous sections between parity-even and parity-odd correlation functions. Until now we considered only exactly conserved currents. We will now consider theories with weakly broken higher-spin symmetry at large $N$. We show that for two-point functions of such currents the relation between even and odd correlators continues to hold. For three-point functions when the spins satisfy triangle inequality, the same relation holds between the even and odd correlators. For weakly broken higher spin currents, parity-odd contribution to correlators exist even when triangle inequality is violated. We show that the relation between even and odd continues to hold in this case.

\subsection*{Three-point function with slightly broken higher spin current}
 In theories with weakly broken higher-spin symmetry even when triangle inequality is violated \eqref{Tinv}, there is one parity-odd contribution in \eqref{sping1}, i.e.
 \begin{align}
     \langle J_{s_1} J_{s_2} J_{s_3}\rangle_{\text{odd}}\ne 0~~~~~~{\rm{even~outside~the~triangle}} . \end{align}
 We also have
 \begin{align}
   &\left\langle J_{s_1}J_{s_2}O_{\Delta} \right\rangle \ne 0 ~~\rm{for ~s_1\ne s_2~\text{for} ~\Delta=1,2}\nonumber\\
   &\left\langle J_{s_1}J_{s_2}O_{\Delta} \right\rangle = \left\langle J_{s_1}J_{s_2}O_{\Delta} \right\rangle_{\textbf{even}}+ \left\langle J_{s_1}J_{s_2}O_{\Delta} \right\rangle_{\text{odd}}
 \end{align} It turns out that all the contribution to the correlator when the spins violate triangle inequality is non-homogeneous \cite{Jain:2021whr}.
 Let us also note that when the spins satisfy triangle inequality, there is no change in the structure of the correlation function.

\subsection*{A derivation of the parity-even-odd relation}
In this section we use slightly broken higher symmetry \cite{Maldacena:2012sf} to derive the relation between the parity-even and the parity-odd part of correlation functions \footnote{In this section we closely follow Appendix D of \cite{Maldacena:2012sf}.}. Let us start with correlation functions of the form $\langle T_{\mu\nu}J_{s_1}J_{s_2}\rangle$ where spins $s_1$ and $s_2$ are arbitrary and the currents need not be conserved. In the following we also use the notation
\begin{align}
    X_{z_i z_j z_k}= X_{\mu\nu\rho} z^{\mu}_{i} z^{\nu}_{j} z^{\rho}_{k} \end{align}
    where $z_i$ are polarization vectors such that 
    \begin{align}
        z_i^2=0,~~~z_i\cdot k_i=0
    \end{align}
    
\subsection{$\langle T_{\mu\nu}J_{s_1}J_{s_2}\rangle$}
Let us consider the spin-4 current $J_4$ that obeys the following non-conservation equation in the quasi-fermionic theory \cite{Maldacena:2012sf}
\begin{equation}
\label{DivJ3AF}
\begin{aligned}
    \partial_{\sigma}J^{\sigma}_{z_1 z_1 z_1} {}& =\frac{80}{7}\frac{\widetilde\lambda}{1+\widetilde\lambda^2} \left(\partial_{z_1}O T_{z_1 z_1}-\frac 25O \partial_{z_1} T_{z_1 z_1}\right)
\end{aligned}
\end{equation}
where we have used the notation $z^{\mu} \partial_{\mu}=\partial_{z} $.
In \eqref{DivJ3AF}, $T$ denotes the trace of the stress tensor \footnote{Even though the stress-tensor is traceless, it can lead to a non-trivial trace WT identity.}, $O$ denotes the scalar operator with scaling dimension $\Delta=2+{\mathcal O}(\frac{1}{N})$ and   $\widetilde\lambda$ is the coupling constant introduced in \cite{Maldacena:2012sf}. 

The charge associated to $J_4$ labelled $Q_4$  has the following action on $O$ 
\begin{equation}
\label{QFQ3algebra}
    \big[ Q_{z_1 z_1 z_1} , O \big] =  \epsilon_{z_1 a b }\partial_{a} \partial_{ z_1} T_{z_1}^{b} +\partial^3_{z_1} O
\end{equation}
Let us now consider the action of $Q_4$ on the three-point correlator $\langle OJ_{s_1}J_{s_2}\rangle$. It leads to the following higher spin equation in position space 
\begin{align}
&\langle [Q_{z_1 z_1 z_1}, O(x_1)]J_{s_1}(x_2)J_{s_2}(x_3)\rangle_\text{QF}+\langle  O(x_1)[Q_{z_1 z_1 z_1}, J_{s_1}(x_2)]J_{s_2}(x_3)\rangle_\text{QF}\cr
&\hspace{.5cm}+\langle O(x_1)J_{s_1}(x_2)[Q_{z_1 z_1 z_1},J_{s_2}(x_3)]\rangle_\text{QF}=\int_x \langle\partial_{\sigma} J^{\sigma}_{z_1 z_1 z_1}(x)O(x_1)\,J_{s_1}(x_2)\,J_{s_2}(x_3)\rangle_\text{QF}
\end{align}
Upon utilizing the algebra in \eqref{QFQ3algebra} and the current equation \eqref{DivJ3AF} we get 
\begin{align}
\label{Q4onJ000QF}
\epsilon_{z_1 ab}\langle\partial_a\partial_{z_1} T^{b}_{z_1}J_{s_1}J_{s_2}\rangle+\langle\text{standard\,terms}\rangle=16\frac{\widetilde\lambda}{1+\widetilde\lambda^2} \int_x \langle O \partial_{z_1} T_{z_1 z_1}OJ_{s_1}J_{s_2}\rangle
\end{align}
where by $\langle\text{standard~terms}\rangle$ we mean correlators with a single insertion of the scalar operator $O$. After a large $N$ factorisation we obtain the following
\begin{align}
\label{Q4onJ000QFPS}
&\epsilon_{z_1 ab}\langle\partial_a\partial_{z_1} T^{b}_{z_1}J_{s_1}J_{s_2}\rangle +\langle\text{standard\,terms}\rangle
=16\frac{\widetilde\lambda}{1+\widetilde\lambda^2} \int_x \langle O(x)O(x_1)\rangle \langle\partial_{z_1} T_{z_1 z_1}(x)J_{s_1}J_{s_2}\rangle
\end{align}
We now express the three-point functions in the quasi-fermionic theory as \cite{Maldacena:2012sf}
\begin{align}
\langle OO\rangle_{\text{QF}}&=(1+\widetilde\lambda^2)\langle OO\rangle_{\text{FF}}\cr
\langle  TJ_{s_1}J_{s_2}\rangle_{\text{QF}}&=\frac{\widetilde\lambda^2}{1+\widetilde\lambda^2}\langle  TJ_{s_1}J_{s_2}\rangle_{\text{FB}}+\frac{1}{1+\widetilde\lambda^2}\langle  TJ_{s_1}J_{s_2}\rangle_{\text{FF}}+\frac{\widetilde\lambda}{1+\widetilde\lambda^2}\langle  TJ_{s_1}J_{s_2}\rangle_{\text{odd}}\cr
\langle  J_{s_1}J_{s_2}O\rangle_{\text{QF}}&=\langle  J_{s_1}J_{s_2}O\rangle_{\text{FF}}+\widetilde\lambda\langle  J_{s_1}J_{s_2}O\rangle_{\text{odd}}
\end{align}
Let us now look at terms with the pole structure $\frac{1}{1+\widetilde\lambda^2}$. This gives us the following relation 
\begin{align}
\label{js1js2tp1PS}
&\epsilon_{z_1 ab}\partial_{a}\partial_{z_1} \left(\langle T^{b}_{z_1}J_{s_1}J_{s_2}\rangle_{\text{FF}}-\langle  T^{b}_{z_1}J_{s_1}J_{s_2}\rangle_{\text{FB}}\right)=-16\int_x\langle \partial_{z_1}T_{z_1 z_1}(x)J_{s_1}J_{s_2}\rangle_{\text{odd}}\langle O(x)O(x_1)\rangle
\end{align}
Let us now look at terms with the pole structure $\frac{\widetilde\lambda}{1+\widetilde\lambda^2}$. This gives us the following relation 
\begin{align}
\label{js1js2tp2PS}
&\epsilon_{z_1 ab}\partial^{a}\partial_{z_1} \langle T^{b}_{z_1}J_{s_1}J_{s_2}\rangle_{\text{odd}}=16\int_x (\langle\partial_{z_1}T_{z_1 z_1}(x)J_{s_1}J_{s_2}\rangle_{\text{FF}}-\langle\partial_{z_1}T_{z_1 z_1}J_{s_1}J_{s_2}\rangle_{\text{FB}})\langle O(x)O(x_1)\rangle
\end{align}
Now we make use of the Todorov operator \cite{Dobrev:1977qv} along with \eqref{WTmom1} and the fact that the trace WT identity is also identical for the free boson and the free fermion. We obtain the following two equations from the above two 
\begin{align}
\label{js1js2tp1PSn}
&\epsilon_{\mu ab}\left(\langle\partial_{a}\partial_{(\nu}  T^{b}_{\rho)}J_{s_1}J_{s_2}\rangle_{\text{FF}}-\langle  T^{b}_{\rho)}J_{s_1}J_{s_2}\rangle_{\text{FB}}\right)+\mu\leftrightarrow\nu+\mu\leftrightarrow\rho\cr
&=-16\int_x\langle \partial_{(\mu}T_{\nu\rho)}(x)J_{s_1}J_{s_2}\rangle_{\text{odd}}\langle O(x)O(x_1)\rangle
\end{align}
and
\begin{align}
\label{js1js2tp2PS1}
&\epsilon_{\mu ab}\langle\partial_{a}\partial_{(\nu}  T^{b}_{\rho)}J_{s_1}J_{s_2}\rangle_{\text{odd}}+\mu\leftrightarrow\nu+\mu\leftrightarrow\rho\cr
&\hspace{2cm}=16\int_x (\langle\partial_{(\mu}T_{\nu\rho)}(x)J_{s_1}J_{s_2}\rangle_{\text{FF}}-\langle\partial_{(\mu}T_{\nu\rho)}(x)J_{s_1}J_{s_2}\rangle_{\text{FB}})\langle O(x)O(x_1)\rangle
\end{align}
In Fourier space, these relations take the following form 
\begin{align}
\label{js1js2tp1FSn}
&\epsilon_{\mu k_1b}k_{1(\nu}\left(\langle  T^{b}_{\rho)}J_{s_1}J_{s_2}\rangle_{\text{FF}}-\langle  T^{b}_{\rho)}J_{s_1}J_{s_2}\rangle_{\text{FB}}\right)+\mu\leftrightarrow\nu+\mu\leftrightarrow\rho=k_1k_{1(\mu}\langle T_{\nu\rho)}J_{s_1}J_{s_2}\rangle_{\text{odd}}
\end{align}
and
\begin{align}
\label{js1js2tp2}
&\epsilon_{\mu k_1b}k_{1(\nu}\langle  T^{b}_{\rho)}J_{s_1}J_{s_2}\rangle_{\text{odd}}+\mu\leftrightarrow\nu+\mu\leftrightarrow\rho= k_1 k_{1(\mu}\left(\langle T_{\nu\rho)}J_{s_1}J_{s_2}\rangle_{\text{FB}}-\langle T_{\nu\rho)}J_{s_1}J_{s_2}\rangle_{\text{FF}}\right)
\end{align}
We now use $\Pi^{\nu\rho}_{\alpha\beta}(k_1)$ to project the RHS and the LHS of the above equations. Using the following 
\begin{align}
 & \Pi^{\nu\rho}_{\alpha\beta}(k_1)k_{1\nu}=   \Pi^{\nu\rho}_{\alpha\beta}(k_1)k_{1\rho}=0\nonumber\\
& \Pi^{\nu\rho}_{\alpha\beta}(k_1) \left(\langle  T_{\nu\rho}J_{s_1}J_{s_2}\rangle_{\text{FB}}-\langle T_{\nu\rho}J_{s_1}J_{s_2}\rangle_{\text{FF}}\right)=\langle T_{\nu\rho}J_{s_1}J_{s_2}\rangle_{\text{FB}}-\langle T_{\nu\rho}J_{s_1}J_{s_2}\rangle_{\text{FF}}
\end{align}
we obtain
\begin{align}
   & k_{1\mu}\epsilon_{(\nu k_1b}\langle T^{b}_{\rho)}J_{s_1}J_{s_2}\rangle_{\text{odd}} = k_1  k_{1\mu}\left(\langle T_{\nu\rho}J_{s_1}J_{s_2}\rangle_{\text{FB}}-\langle T_{\nu\rho}J_{s_1}J_{s_2}\rangle_{\text{FF}}\right)\nonumber\\
   \implies &\epsilon_{(\nu k_1b}\langle T^{b}_{\rho)}J_{s_1}J_{s_2}\rangle_{\text{odd}} = k_1  \left(\langle T_{\nu\rho}J_{s_1}J_{s_2}\rangle_{\text{FB}}-\langle T_{\nu\rho}J_{s_1}J_{s_2}\rangle_{\text{FF}}\right)
\end{align}
Similarly we also obtain 
\begin{align}
\label{tjs1js2eqn}
    k_1\langle T_{\nu\rho}J_{s_1}J_{s_2}\rangle_{\text{odd}}&=-\epsilon_{k_1b(\nu}(\langle T^b_{\rho)}J_{s_1}J_{s_2}\rangle_{\text{FF}}-\langle T^b_{\rho)}J_{s_1}J_{s_2}\rangle_{\text{FB}})
\end{align}
This relation translates to the following in position space
\begin{align}\label{pso12}
    \int\frac{1}{|x-x_1|^4}\langle T_{\nu\rho}J_{s_1}J_{s_2}\rangle_{\text{odd}}=\partial_{x_{1\mu}}\epsilon_{\mu b(\nu}(\langle T^b_{\rho)}J_{s_1}J_{s_2}\rangle_{\text{FF}}-\langle T^b_{\rho)}J_{s_1}J_{s_2}\rangle_{\text{FB}})
\end{align}
In \eqref{tjs1js2eqn} one can as well take the $k_1$ on the LHS to the RHS and upon converting the resulting equation to position space one obtains
\begin{align}\label{pso12}
    \langle T_{\nu\rho}J_{s_1}J_{s_2}\rangle_{\text{odd}}=\int\frac{1}{|x-x_1|^2}\partial_{x_{\mu}}\epsilon_{\mu b(\nu}(\langle T^b_{\rho)}J_{s_1}J_{s_2}\rangle_{\text{FF}}-\langle T^b_{\rho)}J_{s_1}J_{s_2}\rangle_{\text{FB}})
\end{align}
This is precisely the relation between odd and even correlators that we derived in \eqref{evenoddposspace}.
Note that we could have argued this relation directly from \eqref{js1js2tp2PS1} by writing down the explicit structure in position space and realizing that \eqref{pso12} indeed holds. As a simple example we could have $s_1=s_2=2$  which corresponds to the $\langle TTT\rangle$ case. Even though we have used large-$N$ techniques, here we have derived results that are perfectly valid even at finite $N$ as was discussed in previous sections.

Let us now consider the case when the spins of the operators in $\langle TJ_{s_1}J_{s_2}\rangle$ are such that they violate triangle inequality, i.e. $s_1>s_2+2$. In this case the correlators that appear in the higher spin equation are purely non-homogeneous. However, the above relation between parity-odd and parity-even correlators will continue to hold. Our analysis can be easily extended to  operators with higher spin. See Appendix \ref{hsap} for details. 

Note that one can obtain similar relations by working in the quasi-bosonic theory.

\subsection{$\langle J_{s}TO\rangle$}
Let us consider a correlator of the kind $\langle J_{s}TO\rangle$  such that the spins violate triangle inequality, i.e. $s>2$.
We will now derive the relation between the parity-odd part of $\langle J_{s}TO\rangle$ and the same correlator in free theories where it is parity-even.
To do so we consider the action of $Q_4$ on the three-point correlator $\langle J_{s}OO\rangle$. It leads to the following higher spin equation :
\begin{align}
&\langle [Q_{\mu\nu\rho},J_{s}(x_1)]O(x_2)O(x_3)\rangle_\text{QF}+\langle  J_{s}(x_1)[Q_{\mu\nu\rho}, O(x_2)]O(x_3)\rangle_\text{QF}\cr
&\hspace{.5cm}+\langle  J_{s}(x_1)O(x_2)[Q_{\mu\nu\rho},O(x_3)]\rangle_\text{QF}=\int_x \langle\partial_{\sigma} J^{\sigma}_{\mu\nu\rho}(x)J_{s}(x_1)O(x_2)O(x_3)\rangle_\text{QF}
\end{align}
Following similar steps as above we obtain
%
\begin{align}
\label{Q4onJS00QFFTp2}
\langle J_sT_{\nu\rho}O\rangle_{\text{odd}}=\frac{1}{k_2}\epsilon_{(\nu k_2b}\langle J_s T^b_{\rho)}O\rangle_{\text{FF}}
\end{align}
In position space this relation takes the form 
\begin{align}
\label{Q4onJS00QFFTp2PS}
\langle J_sT_{\nu\rho}O\rangle_{\text{odd}}=\int\frac{1}{|x-x_2|^2}\partial_{x_{\mu}}\epsilon_{(\nu\mu b}\langle J_s T^b_{\rho)}O\rangle_{\text{FF}}
\end{align}
This is precisely the relation we obtained in Section \ref{sectionpositionspace}. 
%
The analysis can also be generalised to $\langle J_{s_1}J_{s_2}O\rangle$ of which a special case is $\langle J_{s}J_{s}O\rangle$. 
\subsection{$\langle TOOO\rangle$}\label{Toooo}
In this section we obtain the relation between the parity-odd and the parity-even parts of four-point functions. Let us  consider $\langle TOOO\rangle$. To obtain this correlator we consider the action of $Q_4$ on the four-point function of scalar operators $\langle OOOO\rangle$. 
Following the steps in the previous sections and making use of the following \cite{Jain:2020puw} (a similar identification of the quasi-fermionic correlator has been made in the position and Mellin spaces in \cite{Li:2019twz}) and \cite{Silva:2021ece} respectively)
\begin{align}
\langle  TOOO\rangle_{\text{QF}}&=(1+\widetilde\lambda^2)\langle  TOOO\rangle_{\text{FF}}+\widetilde\lambda(1+\widetilde\lambda^2)\langle  TOOO\rangle_{\text{CB}}
\end{align}
We note that $\langle TOOO\rangle_{\text{FF}}$ is parity-odd and $\langle TOOO\rangle_{\text{CB}}$ is parity-even and they are homogeneous \footnote{The Ward identity is proportional to the 3-point $\langle OOO\rangle$ which can be set to zero for $O$ with scaling dimension 2.} :
\begin{align}
k_1^\mu\langle T_{\mu\nu}OOO\rangle_{FF,CB}=0
\end{align}
From the higher spin equations we obtain 
\begin{align}
\label{Q4onOOOOQFp1PS}
&\left[\epsilon_{\mu ab}\langle\partial_a\partial_{(\nu} T^{b}_{\rho)}OOO\rangle_{\text{FF}}+(\mu\leftrightarrow\nu)+(\mu\leftrightarrow\rho)\right]+\text{permutations}\cr
&=\int_x\bigg[-16 \langle OO\rangle_{\text{FF}}\langle \partial_{(\mu} T_{\nu\rho)}OOO\rangle_{\text{CB}}+(1\leftrightarrow 2)+(1\leftrightarrow 3)+(1\leftrightarrow 4)\bigg]
\end{align}
and
\begin{align}
\label{Q4onOOOOQFp2PS}
&\left[\epsilon_{\mu ab}\langle\partial_a\partial_{(\nu} T^{b}_{\rho)}OOO\rangle_{\text{CB}}+(\mu\leftrightarrow\nu)+(\mu\leftrightarrow\rho)\right]+\text{permutations}\cr
&=\int_x\bigg[ 16\langle OO\rangle_{\text{FF}}\langle \partial_{(\mu} T_{\nu\rho)}OOO\rangle_{\text{FF}}+(1\leftrightarrow 2)+(1\leftrightarrow 3)+(1\leftrightarrow 4)\bigg]
\end{align}
In Fourier space these equations take the following form 
\begin{align}
\label{Q4onOOOOQFp1MS}
&\left[\epsilon_{\mu k_1b}k_{1(\nu}\langle T^{b}_{\rho)}OOO\rangle_{\text{FF}}+(\mu\leftrightarrow\nu)+(\mu\leftrightarrow\rho)\right]+\text{permutations}\cr
&=16 \langle OO\rangle_{\text{FF}}\,k_{1(\mu}\langle T_{\nu\rho)}OOO\rangle_{\text{CB}}+(1\leftrightarrow 2)+(1\leftrightarrow 3)+(1\leftrightarrow 4)
\end{align}
and
\begin{align}
\label{Q4onOOOOQFp2MS}
&\left[\epsilon_{\mu k_1b}k_{1(\nu}\langle T^{b}_{\rho)}OOO\rangle_{\text{CB}}+(\mu\leftrightarrow\nu)+(\mu\leftrightarrow\rho)\right]+\text{permutations}\cr
&=-16\langle OO\rangle_{\text{FF}}\,k_{1(\mu}\langle  T_{\nu\rho)}OOO\rangle_{\text{FF}}+(1\leftrightarrow 2)+(1\leftrightarrow 3)+(1\leftrightarrow 4)
\end{align}
Thus we obtain a relation between the parity-even and the parity-odd parts of $\langle TOOO\rangle$ \footnote{This relation is weaker than the one we discussed in \eqref{Q4onOOOOQFp2MS} because this holds true only with the permutations taken into account. In position space similar relations were explicitly understood in \cite{Li:2019twz}.}.

\section{Summary and future directions}
\label{9}
In this paper we explicitly showed that for exactly conserved currents, one can relate the parity-odd part of the correlation function to the parity-even homogeneous contribution to the correlator \footnote{There is no analogue of parity-odd non-homogeneous correlator which can be obtained from a parity-even non-homogeneous correlator.}. We wrote down an explicit relation between the two in position space, momentum space and in spinor-helicity variables. The fact that, for exactly conserved currents, there does not exist a parity-odd non-homogeneous contribution can be understood by a simple  analysis of the WT identity. However, for weakly broken higher spin theories, when the spins violate triangle inequality there are only non-homogeneous contributions. It turns out that in such cases, a relation exists between the even and the odd non-homogeneous pieces. 
 
 We explicitly wrote down a relation between the parity-even and the parity-odd parts of amplitudes in one higher dimension. We found that in four dimensions  parity-even interactions arising from non-minimal coupling map to parity-odd  interactions arising from non-minimal coupling. We recover the know fact that gravity naturally splits up the correlation into homogeneous and non-homogeneous part \cite{Giombi:2011rz}. We also briefly discussed the case of four-point functions. 
 
 It would be interesting to generalize our results in the following ways. Firstly, it would be interesting to verify the relation between the parity-even and the parity-odd four-point homogeneous correlation functions. 
 It would be interesting to check if similar relations hold for the non-homogeneous parts of four-point functions. One easy way would be to see if the parity-even and the parity-odd WT identities could be mapped to each other in spinor-helicity variables. It would also be interesting to verify if similar relations exist between the parity-even and the parity-odd homogeneous parts of 4 photon and 4 graviton amplitudes \cite{Chowdhury:2019kaq}. 
 
Another interesting direction would be to explore how the relation between even and odd correlators comes about in perturbation theory when we compute correlation functions in Chern-Simons matter theories \cite{Aharony:2012nh,GurAri:2012is,Inbasekar:2019wdw,Kalloor:2019xjb}. The parity-odd contribution comes from odd loop orders whereas parity-even contribution comes from even loop orders. This implies that the relation between the even and odd parts of correlators gives a  relation between the even and odd loop order calculations. The parity-odd contribution leads to anyonic nature exhibited by Chern-Simons matter theories \cite{Jain:2014nza,Inbasekar:2015tsa,Gandhi:2021gwn}. A related story is that of flux attachment \cite{Karch:2016sxi,Seiberg:2016gmd}. It would be interesting to explore these aspects of Chern-Simons matter theories in the light of the relation between the parity-odd and the parity-even correlators that we obtained in this paper. It would be interesting to find out if similar relation holds at finite temperature. This would lead to a relation between viscosity, conductivity and Hall viscosity and Hall conductivity computed in \cite{Geracie:2015drf,Gur-Ari:2016xff}.

\section*{Acknowledgments}
The work of SJ and RRJ is supported by the Ramanujan Fellowship. We acknowledge our debt to the people of India for their steady support of research in basic sciences. We thank S. D. Chowdhury, L. Janagal, S. Sinha and E. Skvortsov for useful discussions and correspondence. We thank A. Mehta and A. Suresh for collaboration at an early stage of this work.

\appendix

\section{Three-point function in momentum space}\label{mom-rev-1}
In this appendix we give details of the three-point functions used in the main text. For our purpose, it is sufficient to concentrate on the homogeneous part of the correlation function. 
The homogeneous part of any $3$-point function of the form $\langle J_{s_1} J_{s_2}J_{s_3} \rangle$ can be written in terms of a finite number of  building blocks which are given below \cite{Jain:2021vrv}.
\begin{align}
    &Q_{12} = \frac{1}{E^{2}}\left[2\left(\vec{z}_{1} \cdot \vec{k}_{2}\right)\left(\vec{z}_{2} \cdot \vec{k}_{1}\right)+E\left(E-2 k_{3}\right) \vec{z}_{1} \cdot \vec{z}_{2}\right]\\
    &S_{12} = \frac{2}{E^{2}}\left[k_{1} \epsilon^{k_{2} z_{1} z_{2}}-k_{2} \epsilon^{k_{1} z_{1} z_{2}}\right]\\
   &P_{123}=\frac{1}{E^{3}}\left[2\left(\vec{z}_{1} \cdot \vec{k}_{2}\right)\left(\vec{z}_{2} \cdot \vec{k}_{3}\right)\left(\vec{z}_{3} \cdot \vec{k}_{1}\right)+E\left(k_{3}\left(\vec{z}_{1} \cdot \vec{z}_{2}\right)\left(\vec{z}_{3} \cdot \vec{k}_{1}\right)+\text { cyclic }\right)\right]\\
   &R_{123} = \frac{1}{E^3}\left[\left\{(\vec{k}_1 \cdot \vec{z}_3)\left(\epsilon^{k_3 z_1 z_2}k_1-\epsilon^{k_1 z_1 z_2}k_3\right)+(\vec{k}_3 \cdot \vec{z}_2)\left(\epsilon^{k_1 z_1 z_3}k_2-\epsilon^{k_2 z_1 z_3}k_1\right)\right.\right.\nonumber\\[5 pt]
&\hspace{1.5cm}\left.\left.-(\vec{z}_2 \cdot \vec{z}_3)\epsilon^{k_1 k_2 z_1}E+\frac{k_1}{2} \epsilon^{z_1 z_2 z_3}E(E-2k_1)\right\}+\text{cyclic perm}\right]
\end{align}

\subsection*{$s_1+s_2+s_3=2n\;\;(n \in \mathbb{Z})$}
 It was shown in \cite{Jain:2021vrv}, for this class of correlators we only require $Q_{ij}$ and $S_{ij}$. 
 Let us consider $\langle J_{s_1}J_{s_2}J_{s_3} \rangle$ such that $s_1 \geq s_2 \geq s_3$, $s_1 \leq s_2+s_3$.  For this case we have
\begin{align}\label{s1s212}
    \langle J_{s_1} J_{s_2}J_{s_3} \rangle_{\text{even}} &= k_1^{s_1-1}k_2^{s_2-1}k_3^{s_3-1}Q_{12}^{\frac{1}{2}(s_1+s_2-s_3)}Q_{23}^{\frac{1}{2}(s_2+s_3-s_1)}Q_{13}^{\frac{1}{2}(s_1+s_3-s_2)}\nonumber\\[5 pt]
   \langle J_{s_1} J_{s_2}J_{s_3} \rangle_{\text{odd}} &= k_1^{s_1-1}k_2^{s_2-1}k_3^{s_3-1}S_{12}Q_{12}^{\frac{1}{2}(s_1+s_2-s_3-2)}Q_{23}^{\frac{1}{2}(s_2+s_3-s_1)}Q_{13}^{\frac{1}{2}(s_1+s_3-s_2)}\nonumber\\[5 pt]
   &+\text{cyclic perm.}
\end{align}
For the correlator with scalar operators we have
\begin{align}\label{Homgeneric}
   &\langle J_sJ_sO_2\rangle_{\text{even},\bf{h}} = b_{12}^{s-1}Q^s_{12}\\
   &\langle J_sJ_sO_2\rangle_{\text{odd},\bf{h}} = b_{12}^{s-1}S_{12}Q^{s-1}_{12}
\end{align}
where $b_{ij}=k_i k_j$ and $c_{123}=k_1 k_2 k_3$. For scalar operator with  $\Delta=1$ we just need to do a shadow transform to \eqref{Homgeneric}. For generic scalar operator dimension $\Delta$ can also be obtained easily.
Now we use following interesting identities
\begin{align}\label{id11}
    \frac{1}{4}E^2 Q_{ij}^2 &= k_i k_j z_i\cdot z_j  Q_{ij}\nonumber\\
    \frac{1}{4}E^2 Q_{ij} S_{ij} &= k_i k_j z_i\cdot z_j  S_{ij}= -k_j\epsilon^{k_i z_i z_j} Q_{ij}= k_i\epsilon^{k_j z_i z_j} Q_{ij}.
\end{align}
We can easily check that 
\begin{align}\label{degen}
 \frac{1}{4}E^2Q_{ij}^2 -   Q_{ij} k_i k_j (z_i\cdot z_j) &= \frac{1}{E^2}\left((k_2 \cdot z_1)^2 (k_3 \cdot z_2)^2+2(k_1 \cdot k_2)(k_2 \cdot z_1)(k_3 \cdot z_2)(z_1 \cdot z_2)-\frac{J^2}{4}(z_1 \cdot z_2)^2\right)\nonumber\\
 &=0
\end{align} where $J^2=4(k_1^2 k_2^2-(k_1 \cdot k_2)^2$.
The last line of \eqref{degen} is zero because RHS is exactly degeneracy factor \cite{Bzowski:2018fql}. Let us emphasize the fact that 
\begin{align}\label{degen1}
 &\frac{1}{4}E^2Q_{ij}^2 - k_i k_j z_i\cdot z_j  Q_{ij}
 =0\nonumber\\
 &\notimplies \frac{1}{4}E^2Q_{ij}= k_i k_j (z_i \cdot z_j) 
\end{align}
This is because the way the degeneracy works is only with $Q_{ij}^2$. Using this identity we can rewrite \eqref{s1s212} as follows
\begin{align}\label{s123a}
 \langle J_{s_1}J_{s_2}J_{s_3} \rangle_{\text{even}} 
  &= \frac{k_1^{2s_1-3}k_2^{2s_2-3}k_3^{2s_3-3}}{E^{s_1+s_2+s_3-6}} Q_{12} Q_{13} Q_{23} \left(z_1\cdot z_2\right)^{\frac{s_1+s_2-s_3-2}{2}} \left(z_2\cdot z_3\right)^{\frac{-s_1+s_2+s_3-2}{2}} \left(z_1\cdot z_3\right)^{\frac{s_1-s_2+s_3-2}{2}} \nonumber\\
   \langle J_{s_1}J_{s_2}J_{s_3} \rangle_{\text{odd}} 
  &= \frac{k_1^{2s_1-3}k_2^{2s_2-3}k_3^{2s_3-3}}{E^{s_1+s_2+s_3-6}} S_{12} Q_{13} Q_{23} \left(z_1\cdot z_2\right)^{\frac{s_1+s_2-s_3-2}{2}} \left(z_2\cdot z_3\right)^{\frac{-s_1+s_2+s_3-2}{2}} \left(z_1\cdot z_3\right)^{\frac{s_1-s_2+s_3-2}{2}} \cr
  &\hspace{1cm}+\text{cyclic}
\end{align}
Remarkably we can use the following identity
\begin{align}\label{id121}
   \frac{1}{4}E^2 Q_{12} Q_{13} Q_{23}&= k_2 k_3 z_2\cdot z_3 Q_{12} Q_{13} = k_1 k_3 z_1\cdot z_3 Q_{12} Q_{23} = k_1 k_2 z_1\cdot z_2 Q_{13} Q_{23}\nonumber\\
    \frac{1}{4}E^2S_{12} Q_{13} Q_{23}&=k_2 k_3 z_2\cdot z_3 S_{12} Q_{13} =k_1 k_3 z_1\cdot z_3 S_{12} Q_{23}
\end{align}
This turns \eqref{s123a} into
\begin{align}\label{s1s2s3fnl}
    \langle J_{s_1}J_{s_2}J_{s_3} \rangle_{\text{even}} 
     = &\frac{k_1^{2s_1-2}k_2^{2s_2-3}k_3^{2s_3-2}}{E^{s_1+s_2+s_3-4}}(z_1 \cdot z_3)^{\frac{s_1-s_2+s_3}{2}}(z_2 \cdot z_3)^{\frac{-s_1+s_2+s_3-2}{2}}(z_1 \cdot z_2)^{\frac{s_1+s_2-s_3-2}{2}} Q_{12} Q_{23}\nonumber\\
  \langle J_{s_1}J_{s_2}J_{s_3} \rangle_{\text{odd}} 
     = &\frac{k_1^{2s_1-2}k_2^{2s_2-3}k_3^{2s_3-2}}{E^{s_1+s_2+s_3-4}}(z_1 \cdot z_3)^{\frac{s_1-s_2+s_3}{2}}(z_2 \cdot z_3)^{\frac{-s_1+s_2+s_3-2}{2}}(z_1 \cdot z_2)^{\frac{s_1+s_2-s_3-2}{2}} S_{12} Q_{23} 
\end{align}
One can symmetrize above in $1,2,3$ however they are just related by degeneracy \footnote{There is a simple interpretation of the formula in \eqref{s1s2s3fnl}. The factors $Q_{ij}$ and $S_{ij}$ are exactly the same as $\langle J J O_2\rangle_{\text{even}}$ and $\langle J J O_2\rangle_{\text{odd}}$ respectively. In spinor-helicity variables $Q^{-+}=Q^{+-}=S^{-+}=S^{+-}=0$ and the only non-zero components are $++$ and $--$.
Since we want only the $---$ and $+++$ components of the homogeneous piece to be non-zero, we require at least two factors of $Q$ or one factor $Q$ and one factor of $S$. This will ensure that any mixed helicity component is zero. We can then appropriately multiply $z_i\cdot z_j$ to account for the spin.}. One can easily check \eqref{s1s2s3fnl} to be correct by going to spinor-helicity variables and matching it with \eqref{s1s2s3sh}. 
For parity-odd case there exists another useful representation. For this we use following identities
\begin{align}\label{dgodd}
   - \frac{1}{4}E^2S_{12} Q_{13} Q_{23}&= k_2\epsilon^{k_1 z_1 z_2} Q_{23} Q_{13} = k_1 \epsilon^{k_2 z_1 z_2} Q_{23} Q_{13} \nonumber\\
    &= k_2\epsilon^{k_3 z_2 z_3} Q_{12} Q_{13} = k_3\epsilon^{k_2 z_2 z_3} Q_{12} Q_{13} \nonumber\\
    &= k_3\epsilon^{k_1 z_1 z_3} Q_{12} Q_{23} =  k_1\epsilon^{k_3 z_1 z_3} Q_{12} Q_{23}
\end{align}
Using this identities we can write the parity-odd result as follows
\begin{align}\label{s1s2s3fnlod}
    \langle J_{s_1}J_{s_2}J_{s_3} \rangle_{\text{odd}} 
     = &\frac{k_1^{2s_1-3}k_2^{2s_2-2}k_3^{2s_3-3}}{E^{s_1+s_2+s_3-4}}(z_1 \cdot z_3)^{\frac{s_1-s_2+s_3-2}{2}}(z_2 \cdot z_3)^{\frac{-s_1+s_2+s_3-2}{2}}(z_1 \cdot z_2)^{\frac{s_1+s_2-s_3-2}{2}}\epsilon^{k_1 z_1 z_2} Q_{13} Q_{23} \nonumber\\
   =  &\frac{k_1^{2s_1-3}k_2^{2s_2-3}k_3^{2s_3-2}}{E^{s_1+s_2+s_3-4}}(z_1 \cdot z_3)^{\frac{s_1-s_2+s_3-2}{2}}(z_2 \cdot z_3)^{\frac{-s_1+s_2+s_3-2}{2}}(z_1 \cdot z_2)^{\frac{s_1+s_2-s_3-2}{2}}\epsilon^{k_2 z_2 z_3} Q_{13} Q_{12}\nonumber\\
   =  &\frac{k_1^{2s_1-2}k_2^{2s_2-3}k_3^{2s_3-3}}{E^{s_1+s_2+s_3-4}}(z_1 \cdot z_3)^{\frac{s_1-s_2+s_3-2}{2}}(z_2 \cdot z_3)^{\frac{-s_1+s_2+s_3-2}{2}}(z_1 \cdot z_2)^{\frac{s_1+s_2-s_3-2}{2}}\epsilon^{k_3 z_1 z_3} Q_{12} Q_{23}.
\end{align}
We can also write the odd part as
\begin{align}\label{s1s2s3fnlod1}
    \langle J_{s_1}J_{s_2}J_{s_3} \rangle_{\text{odd}} 
     = &\frac{k_1^{2s_1-2}k_2^{2s_2-3}k_3^{2s_3-3}}{E^{s_1+s_2+s_3-4}}(z_1 \cdot z_3)^{\frac{s_1-s_2+s_3-2}{2}}(z_2 \cdot z_3)^{\frac{-s_1+s_2+s_3-2}{2}}(z_1 \cdot z_2)^{\frac{s_1+s_2-s_3-2}{2}}\epsilon^{k_2 z_1 z_2} Q_{13} Q_{23} \nonumber\\
   =  &\frac{k_1^{2s_1-3}k_2^{2s_2-2}k_3^{2s_3-3}}{E^{s_1+s_2+s_3-4}}(z_1 \cdot z_3)^{\frac{s_1-s_2+s_3-2}{2}}(z_2 \cdot z_3)^{\frac{-s_1+s_2+s_3-2}{2}}(z_1 \cdot z_2)^{\frac{s_1+s_2-s_3-2}{2}}\epsilon^{k_3 z_2 z_3} Q_{13} Q_{12}\nonumber\\
   =  &\frac{k_1^{2s_1-3}k_2^{2s_2-3}k_3^{2s_3-2}}{E^{s_1+s_2+s_3-4}}(z_1 \cdot z_3)^{\frac{s_1-s_2+s_3-2}{2}}(z_2 \cdot z_3)^{\frac{-s_1+s_2+s_3-2}{2}}(z_1 \cdot z_2)^{\frac{s_1+s_2-s_3-2}{2}}\epsilon^{k_1 z_1 z_3} Q_{12} Q_{23}.
\end{align}
For the case say $s_1=s,s_2=s, s_3= 0$ we can write \eqref{Homgeneric} as
\begin{align}\label{sso1}
   \langle J_s J_s O_2\rangle_{\text{even}} &= \frac{b_{12}^{2s-2}}{E^{2s-2}}Q_{12} \left(z_1\cdot z_2\right)^{s-1}\nonumber\\
    \langle J_s J_s O_2\rangle_{\text{odd}} & =\frac{b_{12}^{2s-2}
     }{E^{2s-2}}S_{12} \left(z_1\cdot z_2\right)^{s-1}\cr
     &=-\frac{b_{12}^{2s-2}
     }{E^{2s-2}}\frac{\epsilon^{k_1z_1z_2}Q_{12}}{k_1} \left(z_1\cdot z_2\right)^{s-2}\cr
     &=\frac{b_{12}^{2s-2}
     }{E^{2s-2}}\frac{\epsilon^{k_2z_1z_2}Q_{12}}{k_2} \left(z_1\cdot z_2\right)^{s-2}
\end{align}
where we have used \eqref{id11}.
As a concrete check 
let us consider  simple example of $\langle TTO \rangle$. In \cite{Jain:2021qcl}, it was shown that 
\begin{align}
    \langle TTO \rangle_{\text{even},\text{there}} &= \frac{k_1 k_2}{E^4}\left(2(k_2 \cdot z_1)(k_3 \cdot z_2)-(z_1 \cdot z_2)E(E-2k_3)\right)^2
\end{align}
Now using \eqref{sso1} we get
\begin{align}
    \langle TTO \rangle_{\text{even},\text{here}} &= \frac{k_1 k_2}{E^4}\left(2(k_2 \cdot z_1)(k_3 \cdot z_2)-(z_1 \cdot z_2)E(E-2k_3)\right)((z_1 \cdot z_2)k_1 k_2)
\end{align}
If we consider the difference between the two expressions after accounting for a relative constant of $\frac 14$ , we get
\begin{align}
    \langle TTO \rangle_{\text{even},\text{here}}-\frac{1}{4}\langle TTO \rangle_{\text{even},\text{there}} &= (k_2 \cdot z_1)^2 (k_3 \cdot z_2)^2+2(k_1 \cdot k_2)(k_2 \cdot z_1)(k_3 \cdot z_2)(z_1 \cdot z_2)-\frac{J^2}{4}(z_1 \cdot z_2)^2\nonumber\\
    &=0
\end{align}
where we have used  \eqref{degen}.

Let us take another example for $\langle TTT \rangle_{\text{odd}}$ illustration. In \cite{Jain:2021vrv} we derived
\begin{align}\label{oldt}
\langle TTT \rangle_{\text{odd},\text{there}} &= \frac{k_1 k_2 k_3}{E^6} S_{12} Q_{23} Q_{13}.
\end{align}
Using representation \eqref{s1s2s3fnlod} or \eqref{s1s2s3fnlod1} we have
\begin{align}
    \langle TTT \rangle_{\text{odd},\text{here}} &= \frac{k_1 k_2 k_3}{E^2}\epsilon^{k_1 z_1 z_3}k_3 Q_{12} Q_{23}
\end{align}It is easy to check using Schouten identity or spinor-helicity variables that the two expressions are the same up to overall numbers. More precisely one can show
\begin{align}
    \langle TTT \rangle_{\text{odd},\text{here}}+2\langle TTT \rangle_{\text{odd},\text{there}} &= 0.
\end{align}
\subsection*{Relating parity-even and parity-odd}
Now we have all the ingredient to relate parity-even and parity-odd results. Let us write the results in way which makes it evident
\begin{align}
   \langle J_s J_s O_2\rangle_{\text{even}} &= \frac{b_{12}^{2s-2}}{E^{2s-2}}Q_{12} \left(z_1\cdot z_2\right)^{s-1}\nonumber\\
    \langle J_s J_s O_2\rangle_{\text{odd}} 
     &=\frac{b_{12}^{2s-2}}{E^{2s-2}}Q_{12}\frac{\epsilon^{k_2 z_1 z_2}}{k_2} \left(z_1\cdot z_2\right)^{s-2}\nonumber\\
      \langle J_{s_1}J_{s_2}J_{s_3} \rangle_{\text{even}} 
     &= \frac{k_1^{2s_1-2}k_2^{2s_2-2}k_3^{2s_3-3}}{E^{s_1+s_2+s_3-4}}(z_1 \cdot z_3)^{\frac{s_1-s_2+s_3-2}{2}}(z_2 \cdot z_3)^{\frac{-s_1+s_2+s_3-2}{2}}(z_1 \cdot z_2)^{\frac{s_1+s_2-s_3}{2}} Q_{13} Q_{23}\nonumber\\
     \langle J_{s_1}J_{s_2}J_{s_3} \rangle_{\text{odd}} 
     &= \frac{k_1^{2s_1-3}k_2^{2s_2-2}k_3^{2s_3-3}}{E^{s_1+s_2+s_3-4}}(z_1 \cdot z_3)^{\frac{s_1-s_2+s_3-2}{2}}(z_2 \cdot z_3)^{\frac{-s_1+s_2+s_3-2}{2}}(z_1 \cdot z_2)^{\frac{s_1+s_2-s_3-2}{2}}\epsilon^{k_1 z_1 z_2} Q_{13} Q_{23}  \end{align}
 We see simple relation between parity-even and parity-odd results
\begin{align}
    \frac{1}{k_1}\epsilon^{z_1 z_2 k_1}  \frac{\partial}{\partial (z_1\cdot z_2)} : {\rm{even_{\textbf{h}}}}\rightarrow {\rm{Odd}}.
\end{align} 
Let us note that for $s_1\ne 0, s_2\ne 0, s_3\ne 0$ we could have chosen any $z_i\cdot z_j$ and replaced it with either $ \frac{1}{k_i}\epsilon^{z_i z_j k_i}$ or with $ \frac{1}{k_j}\epsilon^{z_i z_j k_j}$ it would have given us the map \eqref{opodev11}. Also note that the map in \eqref{opodev11} is exactly same as the map observed at the level of two-point function \eqref{opodev}.

\subsection*{$s_1+s_2+s_3=2n+1\;\;(n \in \mathbb{Z})$}
For this case, we require $P_{123}$ and $R_{123}$ as well. For example, when $s$ is odd, we have
 \begin{align}
 &\langle J_s J_s J_s\rangle_{\text{even},\bf{h}} = c^{s-1}_{123} P^s_{123}\\
   &\langle J_s J_s J_s\rangle_{\text{odd},\bf{h}} = c^{s-1}_{123} R_{123} P^{s-1}_{123}
   \end{align} where $c_{123}=k_1 k_2 k_3$.
   Now using the fact that 
 \begin{align}
     9P_{123}^2 &= R_{123}^2= \frac{16}{9}Q_{12} Q_{13}Q_{23}\nonumber\\
     \frac{1}{4}E^2 P_{123} Q_{ij}&= k_i k_j z_i\cdot z_j P_{123} ,~~ \frac{1}{4}E^2R_{123} Q_{ij}= k_i k_j z_i\cdot z_j R_{123}\nonumber\\
     \end{align}
     we get
\begin{align}
    &\langle J_s J_s J_s\rangle_{\text{even},\bf{h}} = \frac{c^{2(s-1)}_{123}}{E^{3s-3}} P_{123} \left(z_1\cdot z_2 z_1\cdot z_3 z_2\cdot z_3 \right)^{\frac{s-1}{2}}\\
   &\langle J_s J_s J_s\rangle_{\text{odd},\bf{h}} = \frac{c^{2(s-1)}_{123}}{E^{3s-3}} R_{123} \left(z_1\cdot z_2 z_1\cdot z_3 z_2\cdot z_3 \right)^{\frac{s-1}{2}}
\end{align}     
 Finally when $s_1+s_2+s_3=odd$ for general spins, it is easy to check that the answer is 
 \begin{align}
  &\langle J_{s_1} J_{s_2} J_{s_3}\rangle_{\text{even},\bf{h}} = \frac{k_1^{2(s_1-1)}k_2^{2(s_2-1)}k_3^{2(s_3-1)}}{E^{s_1+s_2+s_3-3}} P_{123} \left(z_1\cdot z_2\right)^{a} \left(z_1\cdot z_3\right)^{b} \left(z_2\cdot z_3 \right)^{c}\\
   &\langle J_{s_1} J_{s_2} J_{s_3}\rangle_{\text{odd},\bf{h}}= \frac{k_1^{2(s_1-1)}k_2^{2(s_2-1)}k_3^{2(s_3-1)}}{E^{s_1+s_2+s_3-3}} R_{123} \left(z_1\cdot z_2\right)^{a} \left(z_1\cdot z_3\right)^{b} \left(z_2\cdot z_3 \right)^{c}  
 \end{align}
 where
 \begin{align}
         a&=\frac{1}{2}(s_1+s_2-s_3-1)\nonumber\\
           b&=\frac{1}{2}(s_1+s_3-s_2-1)\nonumber\\
         c&=\frac{1}{2}(s_2+s_3-s_1-1)
 \end{align}
 One can also check that 
 \begin{align}
  \frac{1}{k_1} \epsilon^{k_1 z_1 z_2}  P_{123}= -k_1 k_2 z_1\cdot z_2    R_{123}
 \end{align}
 Further more $R_{123}$ and $P_{123}$ are related by
 \begin{align}
   \frac{1}{k_i}\epsilon^{\mu k_i z_i}\frac{\partial}{\partial z_i^{\mu}} P_{123}=  R_{123}
 \end{align}
 We see that \eqref{oddab} or \eqref{oddcd} maps parity-even correlator to parity-odd correlator.

\section{Higher-spin correlators}\label{hsap}
Let us consider the correlator $\langle J_4 J_{s_1}J_{s_2}\rangle$. We consider the action of the charge $Q_6$ associated to the spin-6 current $J_6$ on the correlator $\langle OJ_{s_1}J_{s_2}\rangle$. To do so we make use of the following algebra 
\begin{align}
    [Q_{----},O]=\epsilon_{-ab}\partial_{a}\partial_{-}J_{b---}+\epsilon_{-ab}\partial_a\partial_{-}^3T_{b-}+\partial_{-}^5O
\end{align}
and the current equation 
\begin{align}
    \partial_\sigma J^\sigma_{-----}=16\frac{\widetilde\lambda}{1+\widetilde\lambda^2}\left[\partial_{-}J_{----}O+\partial_{-}T_{--}\partial_{-}^2O\right]
\end{align}
The higher spin equation that arises from the action of $Q_6$ on $\langle OJ_{s_1}J_{s_2}\rangle$ is 
\begin{align}
&\langle [Q_{6}, O(x_1)]J_{s_1}(x_2)J_{s_2}(x_3)\rangle_\text{QF}+\langle  O(x_1)[Q_{6}, J_{s_1}(x_2)]J_{s_2}(x_3)\rangle_\text{QF}\cr
&\hspace{.5cm}+\langle O(x_1)J_{s_1}(x_2)[Q_{6},J_{s_2}(x_3)]\rangle_\text{QF}=\int_x \langle\partial_{\sigma} J^{\sigma}_{-----}(x)O(x_1)\,J_{s_1}(x_2)\,J_{s_2}(x_3)\rangle_\text{QF}
\end{align}
This leads to 
\begin{align}
&\epsilon_{-ab}\partial_a\partial_-\langle J_{b---}J_{s_1}J_{s_2}\rangle+\epsilon_{-ab}\partial_a\partial_{-}^3\langle T_{b-}J_{s_1}J_{s_2}\rangle+\text{standard terms}\cr
&=\frac{\widetilde\lambda}{1+\widetilde\lambda^2}\int_x\langle OO\rangle\langle\partial_{-}J_{----}J_{s_1}J_{s_2}\rangle+\langle\partial_{-}^2OO\rangle\langle\partial_{-}T_{--}J_{s_1}J_{s_2}\rangle
\end{align}
One can make use of the higher spin equation \eqref{Q4onJ000QFPS} to get rid of correlators involving the stress-tensor from the above equation. This leaves us with the following
\begin{align}
&\epsilon_{-ab}\partial_a\partial_-\langle J_{b---}J_{s_1}J_{s_2}\rangle+\text{standard terms}=\frac{\widetilde\lambda}{1+\widetilde\lambda^2}\int_x\langle\partial_{-}^2OO\rangle\langle\partial_{-}T_{--}J_{s_1}J_{s_2}\rangle
\end{align}
We now express the three-point correlator
\begin{align}
\langle  J_4J_{s_1}J_{s_2}\rangle_{\text{QF}}&=\frac{\widetilde\lambda^2}{1+\widetilde\lambda^2}\langle  J_4J_{s_1}J_{s_2}\rangle_{\text{FB}}+\frac{1}{1+\widetilde\lambda^2}\langle  J_4J_{s_1}J_{s_2}\rangle_{\text{FF}}+\frac{\widetilde\lambda}{1+\widetilde\lambda^2}\langle  J_4J_{s_1}J_{s_2}\rangle_{\text{odd}}
\end{align}
At $\mathcal O(\frac{1}{1+\widetilde\lambda^2})$ we obtain
\begin{align}
&\epsilon_{-ab}(\langle\partial_a\partial_- J_{b---}J_{s_1}J_{s_2}\rangle_{\text{FB}}-\langle\partial_a\partial_- J_{b---}J_{s_1}J_{s_2}\rangle_{\text{FF}})=\int_x\langle O(x)O(x_1)\rangle\langle\partial_{-}J_{----}(x)J_{s_1}J_{s_2}\rangle_{\text{odd}}
\end{align}
At $\mathcal O(\frac{\widetilde\lambda}{1+\widetilde\lambda^2})$ we obtain
\begin{align}
&\epsilon_{-ab}\langle\partial_a\partial_- J_{b---}J_{s_1}J_{s_2}\rangle_{\text{odd}}=\int_x\langle O(x)O(x_1)\rangle(\langle\partial_{-}J_{----}(x)J_{s_1}J_{s_2}\rangle_{\text{FF}}-\langle\partial_{-}J_{----}(x)J_{s_1}J_{s_2}\rangle_{\text{FB}})
\end{align}
In Fourier space one obtains the following equations
\begin{align}
\langle J_{----}J_{s_1}J_{s_2}\rangle_{\text{odd}}=& k_1\epsilon_{-k_1b}(\langle J_{b---}J_{s_1}J_{s_2}\rangle_{\text{FB}}-\langle J_{b---}J_{s_1}J_{s_2}\rangle_{\text{FF}})
\end{align}
and
\begin{align}
& k_1\epsilon_{-k_1b}\langle J_{b---}J_{s_1}J_{s_2}\rangle_{\text{odd}}=\langle J_{----}J_{s_1}J_{s_2}\rangle_{\text{FF}}-\langle J_{----}J_{s_1}J_{s_2}\rangle_{\text{FB}}
\end{align}

\providecommand{\href}[2]{#2}\begingroup\raggedright
\bibliography{refs}
\bibliographystyle{JHEP}
\endgroup

\end{document}